\documentclass[aps,showpacs,twocolumn,superscriptaddress]{revtex4} 
 
\usepackage{graphicx} 
\usepackage{dcolumn} 
\usepackage{amsmath}

\newcommand{\sfrac}[2]{\mbox{\footnotesize $\displaystyle \frac{#1}{#2}$}} 
 
\begin{document} 
 
%\preprint{ANL-PHY-xxxxx-TH-2001} 
 
\title{Bethe-Salpeter equation and a nonperturbative quark-gluon vertex} 
% Optional argument for running titles on pages 
%\title[]{} 
 
 
\author{A.~Bender} 
\affiliation{Special Research Centre for the Subatomic Structure of Matter,\\ 
and Department of Physics and Mathematical Physics, University of Adelaide, 
Adelaide SA 5005, Australia\vspace*{1ex}} 
\author{W.~Detmold} 
\affiliation{Special Research Centre for the Subatomic Structure of Matter,\\ 
and Department of Physics and Mathematical Physics, University of Adelaide, 
Adelaide SA 5005, Australia\vspace*{1ex}} 
\author{C.D.~Roberts} 
%\email[]{cdroberts@anl.gov} 
%\homepage[]{http://www.phy.anl.gov/theory/staff/cdr.html} 
\affiliation{Physics Division, Argonne National Laboratory, Argonne Illinois 
60439-4843\vspace*{1ex}} 
\affiliation{Fachbereich Physik, Universit\"at Rostock, D-18051 Rostock, 
Germany\vspace*{1ex}} 
\author{A.W.~Thomas\vspace*{1ex}} 
\affiliation{Special Research Centre for the Subatomic Structure of Matter,\\ 
and Department of Physics and Mathematical Physics, University of Adelaide, 
Adelaide SA 5005, Australia\vspace*{1ex}} 
%\thanks{} 
 
\begin{abstract} 
\rule{0ex}{3ex} 
A Ward-Takahashi identity preserving Bethe-Salpeter kernel can always be
calculated explicitly from a dressed-quark-gluon vertex whose diagrammatic
content is enumerable.  We illustrate that fact using a vertex obtained via
the complete resummation of dressed-gluon ladders.  While this vertex is
planar, the vertex-consistent kernel is nonplanar and that is true for any
dressed vertex.  In an exemplifying model the rainbow-ladder truncation of
the gap and Bethe-Salpeter equations yields many results; e.g., $\pi$- and
$\rho$-meson masses, that are changed little by including higher-order
corrections. Repulsion generated by nonplanar diagrams in the
vertex-consistent Bethe-Salpeter kernel for quark-quark scattering is
sufficient to guarantee that diquark bound states do not exist.
\end{abstract} 
%\keywords{} 
 
\maketitle 
 
\section{Introduction} 
\label{introduction} 
Dynamical chiral symmetry breaking (DCSB) and confinement are keystones in an
understanding of strong interaction observables and their explanation via a
nonperturbative treatment of QCD.  The gap equation~\cite{fn:Eucl}
\begin{eqnarray} 
\nonumber 
\lefteqn{S(p)^{-1}  =  Z_2 \,(i\gamma\cdot p + m_{\rm bm})}\\ 
&&  +\, Z_1 \int^\Lambda_q \, g^2 D_{\mu\nu}(p-q) \frac{\lambda^a}{2}\gamma_\mu 
S(q) \Gamma^a_\nu(q,p) \,, \label{gendse} 
\end{eqnarray} 
is an insightful tool that has long been used to explore the connection
between these phenomena and the long-range behaviour of the interaction in
QCD~\cite{cdragw}.  In this equation: $D_{\mu\nu}(k)$ is the renormalised
dressed-gluon propagator; $\Gamma^a_\nu(q;p)$ is the renormalised
dressed-quark-gluon vertex; $m_{\rm bm}$ is the $\Lambda$-dependent
current-quark bare mass that appears in the Lagrangian; and $\int^\Lambda_q
:= \int^\Lambda d^4 q/(2\pi)^4$ represents a translationally-invariant
regularisation of the integral, with $\Lambda$ the regularisation mass-scale.
The quark-gluon-vertex and quark wave function renormalisation constants,
$Z_1(\zeta^2,\Lambda^2)$ and $Z_2(\zeta^2,\Lambda^2)$ respectively, depend on
the renormalisation point and the regularisation mass-scale.
 
The solution of Eq.~(\ref{gendse}) is the dressed-quark propagator, which
takes the form
\begin{eqnarray} 
\nonumber 
 S(p)^{-1} & = & i \gamma\cdot p \, A(p^2,\zeta^2) + B(p^2,\zeta^2) \\ 
& =& \frac{1}{Z(p^2,\zeta^2)}\left[ i\gamma\cdot p + M(p^2,\zeta^2)\right] 
\,, \label{sinvp} 
\end{eqnarray} 
and is obtained by solving the gap equation subject to the renormalisation
condition that at some large, spacelike $\zeta^2$
\begin{equation} 
\label{renormS} \left.S(p)^{-1}\right|_{p^2=\zeta^2} = i\gamma\cdot p + 
m(\zeta)\,, 
\end{equation} 
where $m(\zeta)$ is the renormalised current-quark mass at the scale $\zeta$: 
$Z_4\,m(\zeta) = Z_2 \, m_{\rm bm}$, with $Z_4$ the renormalisation constant 
for the scalar part of the quark self energy.  At one loop in perturbation 
theory 
\begin{equation} 
m(\zeta) = \frac{\hat m}{\left( \ln [ \zeta/\Lambda_{\rm QCD}] 
\right)^{\gamma_m}} \,, 
\end{equation} 
where: $\hat m$ is the renormalisation-group-invariant current-quark mass; 
$\gamma_m=12/(33- 2 N_f)$ is the leading-order mass anomalous dimension, for 
$N_f$ active flavours; and $\Lambda_{\rm QCD}$ is the $N_f$-flavour QCD 
mass-scale. 
 
Since QCD is an asymptotically free theory the chiral limit is unambiguously 
defined by $\hat m=0$~\cite{mrt98}, which can be implemented in 
Eq.~(\ref{gendse}) by applying~\cite{mr97} 
\begin{equation} 
\label{chirallimit} \lim_{\Lambda\to \infty} Z_2(\zeta^2,\Lambda^2) \, 
m(\Lambda) = 0\,. 
\end{equation} 
The formation of a \textit{gap}, described by Eq.~(\ref{gendse}), is
identified with the appearance of a solution for the dressed-quark propagator
in which $m(\zeta)\sim {\rm O}(1/\zeta^2)\neq 0$; i.e., a solution in which
the mass function is power-law suppressed.  This is DCSB.  It is impossible
at any finite order in perturbation theory and entails the appearance of a
nonzero value for the vacuum quark condensate:
\begin{equation} 
\label{qbq0} \,-\,\langle \bar q q \rangle_\zeta^0 = \lim_{\Lambda\to \infty} 
Z_4(\zeta^2,\Lambda^2)\, N_c \, {\rm tr}_{\rm D}\int^\Lambda_q\!
S^{0}(q,\zeta)\,,  
\end{equation} 
where ${\rm tr}_D$ identifies a trace over Dirac indices alone and the 
superscript ``$0$'' indicates the quantity was calculated in the chiral limit. 
 
It is apparent that the kernel of the gap equation is formed from a product of 
the dressed-gluon propagator and dressed-quark-gluon vertex.  The kernel may be 
calculated in perturbation theory but that is inadequate for the study of 
intrinsically nonperturbative phenomena.  Consequently, to make 
model-independent statements about DCSB one must employ an alternative 
systematic and chiral symmetry preserving truncation scheme. 
 
One such scheme was introduced in Ref.~\cite{truncscheme}.  Its leading-order
term is the rainbow-ladder truncation of the DSEs and the general procedure
provides a means to identify, \textit{a priori}, those channels in which that
truncation is likely to be accurate.  This scheme underlies the successful
application of a renormalisation-group-improved rainbow-ladder model to
flavour-nonsinglet pseudoscalar mesons~\cite{mr97} and vector
mesons~\cite{pieterrho,pieterpion,pieterpiK,pieterother}, and indicates why
the leading-order truncation is inadequate for scalar mesons and
flavour-singlet pseudoscalars~\cite{cdrqcII}. The systematic nature of the
scheme has also made possible a proof of Goldstone's theorem in
QCD~\cite{mrt98}.
 
In quantitative applications, however, the leading-order term alone has been
used almost exclusively: Refs.~\cite{pieterpiK,arne} are exceptions but
they consider just the next-to-leading-order term. Hence one goal of our
study is a nonperturbative verification of the leading-order truncation's
accuracy.
 
One element of the gap equation's kernel is the dressed-gluon propagator, which 
in Landau gauge can be written 
\begin{equation} 
D_{\mu\nu}(k) = \left[ \delta_{\mu\nu} - \frac{k_\mu k_\nu}{k^2} \right] 
\frac{d(k^2,\zeta^2)}{k^2}\,. 
\end{equation} 
It has been the focus of DSE studies~\cite{ragluon} and lattice
simulations~\cite{tonylattice,kurtlattice}, and contemporary analyses suggest
that $d(k^2,\zeta^2)/k^2$ is finite, and of O$(1\,{\rm GeV}^{-2})$, at
$k^2=0$.  However, this behaviour is difficult to reconcile with the
existence and magnitude of DCSB in the strong interaction
spectrum~\cite{fredIR}: it is a model-independent result that a description
of observable phenomena requires a kernel in the gap equation with
significant integrated strength on the domain $k^2\lesssim
1\,$GeV$^2$~\cite{cdresi}.  The required magnification may arise via an
enhancement in the dressed-quark-gluon vertex but, hitherto, no calculation
of the vertex exhibits such behaviour~\cite{latticevertex}.  Hence another
aim of our study is to contribute to the store of nonperturbative information
about this vertex.
 
In Sec.~\ref{SecTwo} we briefly recapitulate on the truncation scheme of
Ref.~\cite{truncscheme}.  Then, using a simple confining model~\cite{mn83},
we demonstrate that an infinite subclass of contributions to the
dressed-quark-gluon vertex: the dressed-gluon ladders, can be resummed via an
algebraic recursion relation, which provides a closed form result for the
vertex expressed solely in terms of the dressed-quark propagator. This
facilitates a simultaneous solution of the coupled gap and vertex equations
obtained via the infinite resummation, as we describe in Sec.~\ref{SecThree}.
While the algebraic simplicity of these results is peculiar to our
rudimentary model, we anticipate that the qualitative behaviour of the
solutions is not.
 
In Sec.~\ref{SecFour} we describe the general procedure that enables a
calculation of the Bethe-Salpeter kernel for flavour-nonsinglet mesons that
is consistent with the fully-resummed dressed-gluon-ladder vertex. The kernel
is itself a resummation of infinitely many diagrams; and it is \textit{not}
planar, an outcome necessary to ensure the preservation of Ward-Takahashi
identities.  This kernel is the heart of the inhomogeneous vertex equations
and associated bound state equations whose solutions relate to strong
interaction observables.  In our simple model it, like the vertex, can also
be obtained via an algebraic recursion relation, complete and in a practical
closed form.  In Sec.~\ref{SecFour} we also study the bound state equations
in a number of meson channels, and derive and solve the analogous equation
for diquark channels.
 
Section~\ref{SecFive} is a summary. 
 
\section{A Dressed-Quark-Gluon Vertex} 
\label{SecTwo} 
The truncation scheme introduced in Ref.~\cite{truncscheme} may be described
as a dressed-loop expansion of the dressed-quark-gluon vertices that appear
in the half amputated dressed-quark-antiquark (or -quark-quark) scattering
matrix: $S^2 K$, a renormalisation-group invariant, where $K$ is the
dressed-quark-antiquark (or -quark-quark) scattering kernel. All $n$-point
functions involved thereafter in connecting two particular quark-gluon
vertices are \textit{fully dressed}.  The effect of this truncation in the
gap equation, Eq.~(\ref{gendse}), is realised through the following
representation of the dressed-quark-gluon vertex, $i \Gamma_\mu^a =
\sfrac{i}{2}\lambda^a\,\Gamma_\mu = l^a \Gamma_\mu$:
\begin{eqnarray} 
\nonumber 
\lefteqn{ Z_1 \Gamma_\mu(k,p)  =  \gamma_\mu }\\ 
\nonumber &+ &  \sfrac{1}{2 N_c} \int_\ell^\Lambda\! g^2 D_{\rho\sigma}(p-\ell) 
\gamma_\rho S(\ell+k-p) \gamma_\mu S(\ell) 
\gamma_\sigma\\ 
\nonumber &+ & \sfrac{N_c}{2}\int_\ell^\Lambda\! g^2\, 
D_{\sigma^\prime \sigma}(\ell) \, D_{\tau^\prime\tau}(\ell+k-p)\, \\ 
 & & \times \, \gamma_{\tau^\prime} \, S(p-\ell)\, 
\gamma_{\sigma^\prime}\, 
\Gamma^{3g}_{\sigma\tau\mu}(\ell,-k,k-p) + [\ldots]\,. 
\label{vtxexpand} 
\end{eqnarray} 
Here $\Gamma^{3g}$ is the dressed-three-gluon vertex and it is apparent that
the lowest order contribution to each term written explicitly is O$(g^2)$.
The ellipsis represents terms whose leading contribution is O$(g^4)$; i.e.,
the crossed-box and two-rung dressed-gluon ladder diagrams, and also terms of
higher leading-order.
 
The expansion of $S^2 K$ just described, with its implications for other
$n$-point functions; e.g., the dressed-quark-photon vertex, yields an ordered
truncation of the DSEs that, term-by-term, guarantees the preservation of
vector and axial-vector Ward-Takahashi identities, a feature exploited in
Ref.~\cite{mrt98} to prove Goldstone's theorem. Furthermore, it is readily
seen that inserting Eq.~(\ref{vtxexpand}) into Eq.~(\ref{gendse}) provides
the rule by which the renormalisation-group-improved rainbow-ladder
truncation~\cite{mr97,pieterrho,pieterpion,pieterpiK,pieterother} can be
systematically improved.  It thereby facilitates an explicit enumeration of
corrections to the impulse current that is widely used in calculations of
electroweak hadron form factors~\cite{pieterpiK}.
 
\subsection{Resumming Dressed-Gluon Ladders} 
\begin{figure}[t] 
 
\centerline{\includegraphics[height=8em]{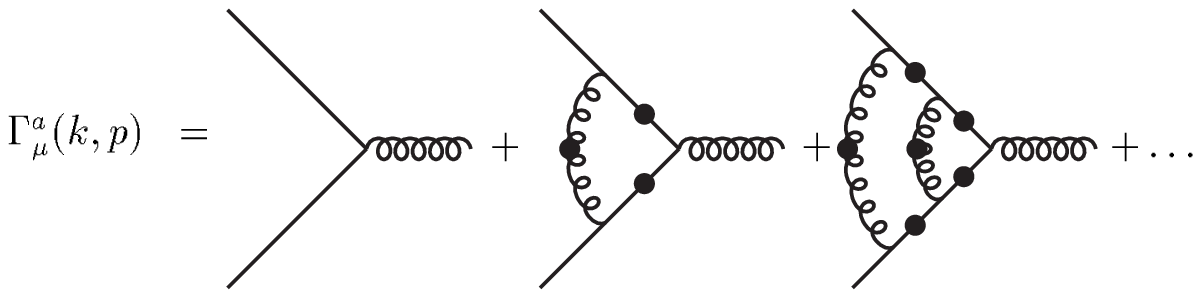}}
 
\caption{\label{Gamma_inf} Integral equation for a planar dressed-quark-gluon
vertex obtained by neglecting contributions associated with explicit gluon
self-interactions.  Solid circles indicate fully dressed propagators.  The
vertices are not dressed.}
\end{figure}

We can't say anything about a complete resummation of the terms in 
Eq.~(\ref{vtxexpand}). However, we are able to contribute to aspects of the 
more modest problem obtained by retaining only the sum of dressed-gluon 
ladders; i.e., aspects of the vertex depicted in Fig.~\ref{Gamma_inf}.  This 
infinite subclass of diagrams is $1/N_c$-suppressed. 

\subsubsection{A Model} 
To simplify our analysis and make the key elements transparent we employ the
confining model introduced in Ref.~\cite{mn83}, which is defined by the
following choice for the dressed-gluon line in Fig.~\ref{Gamma_inf}:
\begin{equation} 
\label{mnmodel} {\cal D}_{\mu\nu}(k):= g^2 \, D_{\mu\nu}(k) = 
\left(\delta_{\mu\nu} - \frac{k_\mu k_\nu}{k^2}\right) (2\pi)^4\, {\cal G}^2 \, 
\delta^4(k)\,. 
\end{equation} 
Plainly, ${\cal G}$, measured in GeV, sets the model's mass-scale and
henceforth we set ${\cal G}=1$ so that all mass-dimensioned quantities are
measured in units of ${\cal G}$.  In the following, since the model is
ultraviolet-finite, we usually remove the regularisation mass-scale to
infinity and set the renormalisation constants equal to one.
 
The model defined by Eq.~(\ref{mnmodel}) is a precursor to an efficacious
class of models that employ a renormalisation-group-improved effective
interaction and whose contemporary application is reviewed in
Refs.~\cite{revbasti,revreinhard}.  It has many positive features in common
with that class and, furthermore, its particular momentum-dependence works to
advantage in reducing integral equations to algebraic equations that preserve
the character of the original equation.  Naturally, there is a drawback: the
simple momentum dependence also leads to some model-dependent artefacts, but
they are easily identified and hence are not cause for concern.
 
\subsubsection{Planar Vertex} 
The general form of the dressed-quark gluon vertex involves twelve distinct
scalar form factors but using Eq.~(\ref{mnmodel}) that part of this vertex
which contributes to the gap equation has no dependence on the total momentum
of the quark-antiquark pair; i.e., only $\Gamma_\mu(p) :=\Gamma_\mu(p,p)$
contributes. This considerably simplifies the analysis since, in general, one
can write
\begin{eqnarray} 
\nonumber \Gamma_\mu(p) & = & \alpha_1(p^2)\, \gamma_\mu
+ \,  \alpha_2(p^2)\, \gamma\cdot p\,p_\mu \\ 
\label{vtxmn} &&   - \, \alpha_3(p^2)\, i \,p_\mu 
+ \alpha_4(p^2) \, i \gamma_\mu \,\gamma\cdot p\,,
\end{eqnarray} 
but it does not restrict our ability to address the questions we raised in
the introduction because those amplitudes which survive are the most
significant in the dressed-quark-photon vertex~\cite{pieterpion} and it is an
enhancement in the vicinity of $p^2=0$ that may be important for a
realisation of DCSB using an infrared-finite dressed-gluon
propagator~\cite{fredIR}.
 
The summation depicted in Fig.~\ref{Gamma_inf} is expressed via
\begin{eqnarray} 
\nonumber \lefteqn{ \Gamma_\mu(p_+,p_-) = Z_1^{-1}\gamma_\mu }\\ 
&& \nonumber   + 
\,\sfrac{1}{6} \int_\ell^\Lambda\! \,{\cal D}_{\rho\sigma}(p-\ell)
\gamma_\rho\, S(\ell_+)\,
\Gamma_\mu(\ell_+,\ell_-)\, S(\ell_-)\, \gamma_\sigma \\
&& \label{GammainfEq} 
\end{eqnarray} 
with $v_\pm = v \pm \sfrac{1}{2}P$, where $v=\ell,p$, etc., is any
four-vector, but using Eq.~(\ref{mnmodel}) this simplifies to
\begin{equation} 
\label{vtxalgebraic} \Gamma_\mu(p) = \gamma_\mu + \sfrac{1}{8}\,\gamma_\rho\, 
S(p)\, \Gamma_\mu(p)\, S(p)\, \gamma_\rho\,, 
\end{equation} 
where the additional factor of $3/4$ ($1/8$ cf.\ $1/6$) owes itself to the
combined operation of the $\delta$-function and the longitudinal projection
operator.  Inserting Eq.~(\ref{vtxmn}) into Eq.~(\ref{vtxalgebraic}) one
finds $ \alpha_4\equiv 0$ and hence the solution of Eq.~(\ref{vtxalgebraic})
simplifies:
\begin{equation} 
\label{vtxinfty} \Gamma_\mu(p) = \alpha_1(p^2)\, \gamma_\mu + \, 
\alpha_2(p^2)\, \gamma\cdot p\,p_\mu - \, \alpha_3(p^2)\, i \,p_\mu\,. 
\end{equation} 
 
We re-express this vertex as 
\begin{eqnarray} 
\label{vtxsummeda} 
\lefteqn{\Gamma_\mu(p) = \sum_{i=0}^\infty\,\Gamma_\mu^i(p)} \\ 
\nonumber &=&  \sum_{i=0}^\infty\, \left[ \alpha^i_1(p^2)\, \gamma_\mu + \, 
\alpha^i_2(p^2)\, \gamma\cdot p\,p_\mu - \, \alpha^i_3(p^2)\, i 
\,p_\mu\right],\\ 
&& \label{vtxi} 
\end{eqnarray} 
where the superscript enumerates the order of the iterate: $\Gamma_\mu^{i=0}$
is the bare vertex,
\begin{equation} 
\begin{array}{cc} 
\alpha_1^0 = 1\,,\; & \alpha_2^0=0=\alpha_3^0\,; 
\end{array} 
\end{equation} 
$\Gamma_\mu^{i=1}$ is the result of inserting this into the r.h.s.\ of
Eq~(\ref{vtxalgebraic}) to obtain the one-rung dressed-gluon correction;
$\Gamma_\mu^{i=2}$ is the result of inserting $\Gamma_\mu^{i=1}$, and is
therefore the two-rung dressed-gluon correction; etc.
 
\begin{figure}[t] 
 
\centerline{\includegraphics[height=12em]{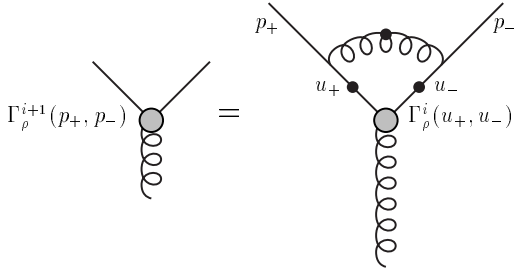}}
 
\caption{\label{gamma_recursion} Recursion relation for the iterates in the
fully-resummed dressed-gluon-ladder vertex,
Eq.~(\protect\ref{vtxrecursfull}): filled circles denote a fully-dressed
propagator or vertex.  Using Eq.~(\protect\ref{mnmodel}), $p=k$, and this
relation is expressed by Eq.~(\protect\ref{vtxrecurs}).}
\end{figure} 
 
Now a simple but important observation is that each iterate is related to its
precursor via the following recursion relation:
\begin{eqnarray} 
\nonumber\lefteqn{ \Gamma_\rho^{a;i+1}(p_+,p_-) := l^a\, 
\Gamma_\rho^{i+1}(p_+,p_-)  = }\\ 
\nonumber && \int_u^\Lambda\! {\cal D}_{\mu\nu}(p-u) \, l^b \gamma_\mu 
\, S(u_+) \, l^a \Gamma_\rho^{i}(u_+,u_-) \, S(u_- )\, l^b 
\gamma_\nu \,, \\
& & \label{vtxrecursfull}
\end{eqnarray} 
which is depicted in Fig.~\ref{gamma_recursion}.  Using
Eq.~(\protect\ref{mnmodel}) this simplifies:
\begin{equation} 
\label{vtxrecurs} \Gamma_\rho^{i+1}(p) = \sfrac{1}{8}\,\gamma_\mu\, S(p)\, 
\Gamma_\rho^i(p)\, S(p)\, \gamma_\mu
\end{equation} 
and substituting Eq.~(\ref{vtxi}) into Eq.~(\ref{vtxrecurs}) yields ($s=p^2$)
\begin{equation} 
\label{matrixrecurs} \mbox{\boldmath $\alpha$}^{i+1}(s):= 
\left( 
\begin{array}{l} 
\rule{0ex}{2.5ex}\alpha_1^{i+1}(s) \\\rule{0ex}{2.5ex} \alpha_2^{i+1}(s) \\ 
\rule{0ex}{2.5ex}\alpha_3^{i+1}(s) 
\end{array}\right) 
= {\cal O}(s;A,B)\, \mbox{\boldmath $\alpha$}^{i}(s)\,, 
\end{equation} 
\begin{eqnarray} 
\nonumber \lefteqn{{\cal O}(s;A,B) = \sfrac{1}{4} \,\frac{1}{(s A^2 + 
B^2)^2}}\\ 
\nonumber && 
\times \left( 
\begin{array}{ccc} 
\rule{0ex}{2.5ex} - (s A^2 + B^2) & 0 & 0\\ 
\rule{0ex}{2.5ex} 2  A^2 & s A^2 - B^2 & 2 A B \\ 
\rule{0ex}{2.5ex} 4 A B & 4 s A B & 2 (B^2 - s A^2) 
\end{array} \right). \\ 
\end{eqnarray} 
Now it is clear that Eqs.~(\ref{vtxinfty}), (\ref{vtxi}), (\ref{matrixrecurs}) 
entail 
\begin{equation} 
\mbox{\boldmath $\alpha$} 
= \left(\sum_{i=1}^\infty\, {\cal O}^i\right) \, \mbox{\boldmath $\alpha$}^0 
= \frac{1}{1 - {\cal O}} \,\mbox{\boldmath $\alpha$}^0\,, 
\end{equation} 
where the last step is valid whenever an iterative solution of 
Eq.~(\ref{GammainfEq}) exists, and defines a solution otherwise, so that, with 
$\Delta = s A^2(s) + B^2(s)$, 
\begin{eqnarray} 
\nonumber 
\alpha_1 & = & \frac{4\, \Delta}{1 + 4\,\Delta}\,,\\ 
 \alpha_2 & = & \frac{- \,8 \,A^2} { 
1 + 2\,( B^2 - s\,A^2) - 8\,\Delta^2 } \, 
\frac{1 + 2\,\Delta}{1 + 4\,\Delta} \,, \label{alpharesults}\\ 
\nonumber \alpha_3 & = & 
\frac{- 8 \,A B}{ 1 + 2 (B^2 - s\, A^2) - 8\,\Delta^2 }\,. 
\end{eqnarray} 
 
We have thus arrived at a closed form for the gluon-ladder-dressed
quark-gluon vertex of Fig.~\ref{Gamma_inf}; i.e., Eqs.~(\ref{vtxinfty}),
(\ref{alpharesults}). Its momentum-dependence is determined by that of the
dressed-quark propagator, which is obtained by solving the gap equation,
itself constructed with this vertex.  Using Eq.~(\ref{mnmodel}) that gap
equation is
\begin{equation} 
S(p)^{-1} = 
\left\{ 
\begin{array}{l} 
 i \gamma\cdot p + m + \gamma_\mu \, S(p) \,\Gamma_\mu(p) \\ 
\rule{0ex}{3ex} i \gamma\cdot p + m + \Gamma_\mu(p) \, S(p)\, \gamma_\mu 
\end{array}\right. 
\label{gapmodel} 
\end{equation} 
and substituting Eq.~(\ref{vtxinfty}) this gives
\begin{eqnarray} 
\nonumber A(s) & = & 1 +\frac{1}{s A^2 + B^2} \left[ A \, (2 \alpha_1 - s 
\alpha_2) - B \,\alpha_3\right]\,,\\ 
&& \label{Afull}\\ 
\nonumber B(s) & = & m+ \frac{1}{s A^2 + B^2} \left[ 
B \, ( 4 \alpha_1 + s \alpha_2) - s A \,\alpha_3 \right]\,. \\ 
\label{Bfull} 
\end{eqnarray} 
 
Obviously, Eqs.~(\ref{Afull}), (\ref{Bfull}), completed using
Eqs.~(\ref{alpharesults}), form a closed, algebraic system.  It can easily be
solved numerically, and that procedure yields simultaneously the complete
gluon-ladder-dressed vertex and the propagator for a quark fully dressed via
gluons coupling through this nonperturbative vertex.
 
We note here that in the chiral limit a realisation of chiral symmetry in the
Wigner-Weyl mode is always possible. This realisation is expressed via the
$B\equiv 0$ solution of the gap equation, which from
Eqs.~(\ref{alpharesults}), (\ref{Bfull}) is evidently always admissible for
$m=0$.
 
\section{Solutions of the Gap and Vertex Equations} 
\label{SecThree} 
\subsection{Algebraic Results} 
\label{secthreeone} 
Before reporting results obtained via a numerical solution we consider a
special case that signals the magnitude of the effects produced by the
complete resummation in Fig.~\ref{Gamma_inf}; i.e., we focus on the solutions
at $s=0$.  In this instance Eqs.~(\ref{alpharesults}) give, with $A_0= A(0)$,
$B_0= B(0)$,
\begin{eqnarray} 
\nonumber 
\alpha_1(s=0) & = & \displaystyle \frac{4\, B_0^2}{1 + 4\, B_0^2} \,, \\ 
 \alpha_2(s=0) & = & \displaystyle\frac{ 8 \, A_0^2}{(4\, B_0^2 + 
1)^2}\, \frac{2 \,B_0^2 + 
1}{2\,B_0^2 - 1}\,, \label{alpha0}\\ 
\nonumber \alpha_3(s=0) & = & \displaystyle\frac{ 8\,A_0\,B_0}{4\,B_0^2 + 
1}\,\frac{1}{2 \, B_0^2 - 1}\,. 
\end{eqnarray} 
 
Substituting these expressions, Eq.~(\ref{Bfull}) becomes 
\begin{equation} 
\label{B0full} B_0 = m + \frac{16\,B_0}{4\,B_0^2+1}\, 
\end{equation} 
and in the chiral limit this yields 
\begin{equation} 
\label{B0res} B_0 = \sfrac{1}{2}\,\sqrt{15} \approx 1.94\,, 
\end{equation} 
which makes plain that the model specified by Eq.~(\ref{mnmodel}) supports a
realisation of chiral symmetry in the Nambu-Goldstone mode; i.e., DCSB. The
value in Eq.~(\ref{B0res}) can be compared with that obtained using the bare
vertex; i.e., the leading-order term in the truncation of
Ref.~\cite{truncscheme}: $B_0^{(0)} = 2$, to see that the completely resummed
dressed-gluon-ladder vertex alters $B_0$ by only $-3\,$\%.
 
Similarly, Eq.~(\ref{Afull}) becomes 
\begin{equation} 
A_0 = 1 + \frac{8\,A_0}{4 \,B_0^2 + 1}\left[ 1 + \frac{1}{2\,B_0^2-1} 
\right]\,, 
\end{equation} 
which using Eq.~(\ref{B0res}) gives
\begin{equation} 
\label{A0res} A_0 = \frac{26}{15} \approx 1.73\,. 
\end{equation} 
This, too, may be compared with the leading-order result: $A_0^{(0)} = 2$. 
Again the resummation does not materially affect the value: here the change is 
$-13\,$\%. 
 
Inserting Eqs.~(\ref{B0res}), (\ref{A0res}) into Eqs.~(\ref{alpha0}) one
finds
\begin{equation} 
\begin{array}{lcccll} 
\alpha_1(0)  & = & \sfrac{15}{16} & \approx & 0.94 & {\rm cf.}~1.0 \,, \\[2ex] 
\alpha_2(0) &= &\sfrac{221}{1800} &\approx & 0.12 & {\rm cf.}~0.0 \,, \\[2ex] 
\alpha_3(0) &= &\sfrac{1}{\sqrt{15}} &\approx & 0.26 & {\rm cf.}~0.0\,, 
\end{array} 
\end{equation} 
where the last, comparative column lists the values for the leading-order
(bare) vertex. It is evident that the solution of Eq.~(\ref{GammainfEq})
obtained using an infrared amplified effective interaction,
Eq.~(\ref{mnmodel}), which supports DCSB (and also
confinement~\cite{fn:confinement}), does not exhibit an enhancement in a
neighbourhood of $p^2=0$.  This provides an internally consistent picture: a
dressed-quark-antiquark scattering kernel, whose embedding in the gap
equation already possesses sufficient integrable strength, does not
additionally magnify itself.
 
\subsection{Numerical Results} 
\subsubsection{Wigner-Weyl Mode} 
The $B\equiv 0$ solution of Eq.~(\ref{Bfull}) is always admitted when $m=0$.
In that case Eq.~(\ref{Afull}) becomes
\begin{equation} 
\label{AsWW} A(s) = 1 + \frac{8\,A(s)}{1 + 4\,s\,A^2(s)} \left[ 1 + \frac{1}{1 
- 4\,s\,A^2(s)}\right]\,, 
\end{equation} 
of which there is no closed-form solution.  (It is a quintic equation for 
$A(s)$.) 
 
\begin{figure}[t] 

\centerline{\includegraphics[height=16em]{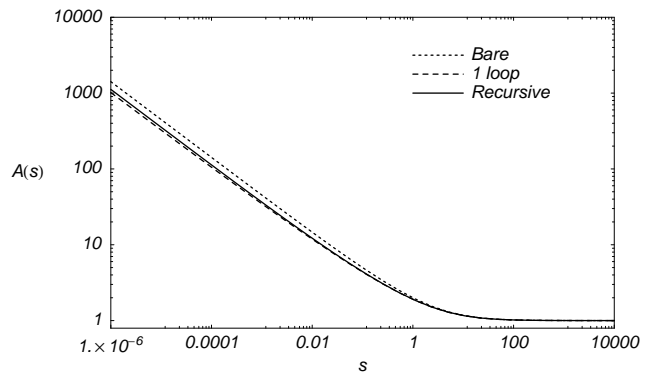}}

\caption{\label{Awwpic} Numerical solution of Eq.~(\ref{AsWW}): solid line; 
cf.\ the solution obtained with the bare vertex: dotted line; and the one-loop 
corrected vertex: dashed line.} 
\end{figure} 
 
However, using the bare vertex one finds a chiral limit solution: $A^{(0)}(s) = 
\sqrt{2/s}$ for $s\approx 0$.  We therefore suppose that in the neighbourhood 
of $s=0$ Eq.~(\ref{AsWW}) admits a solution of the form 
\begin{equation} 
\label{Assmalls} A_{s\sim 0}(s) = \left(\frac{\kappa_A}{s}\right)^{1/2}\,. 
\end{equation} 
Substituting this into Eq.~(\ref{AsWW}) yields 
\begin{equation} 
1 = \frac{16}{1 + 4\,\kappa_A}\,\frac{ 1 - 2\, \kappa_A}{1 - 4\,\kappa_A}\,, 
\end{equation} 
which has two solutions: $\kappa_A = 3/4$, $5/4$.  The required (physical) 
solution of Eq.~(\ref{AsWW}) satisfies $A(s) \stackrel{s\to\infty}{\to} 1^+$. 
Therefore $s A^2(s) \to s A^2_{s\sim 0}$ from above and hence the physical 
branch of the solution is described by 
\begin{equation} 
\kappa_A = \sfrac{5}{4} \, < 2\,. 
\end{equation} 
Evidently, in the chirally symmetric case too, the ladder-dressed vertex 
reduces the magnitude of the solution. 
 
These features are apparent in the complete numerical solution of 
Eq.~(\ref{AsWW}), which is depicted in Fig.~\ref{Awwpic}.  Furthermore, the 
gluon-ladder-dressed vertex, Eq.~(\ref{vtxinfty}), obtained with this solution 
does not exhibit an enhancement. 
 
\subsubsection{Nambu-Goldstone Mode} 
\label{secthreetwo} 
\begin{figure}[t] 
 
\centerline{\includegraphics[height=16em]{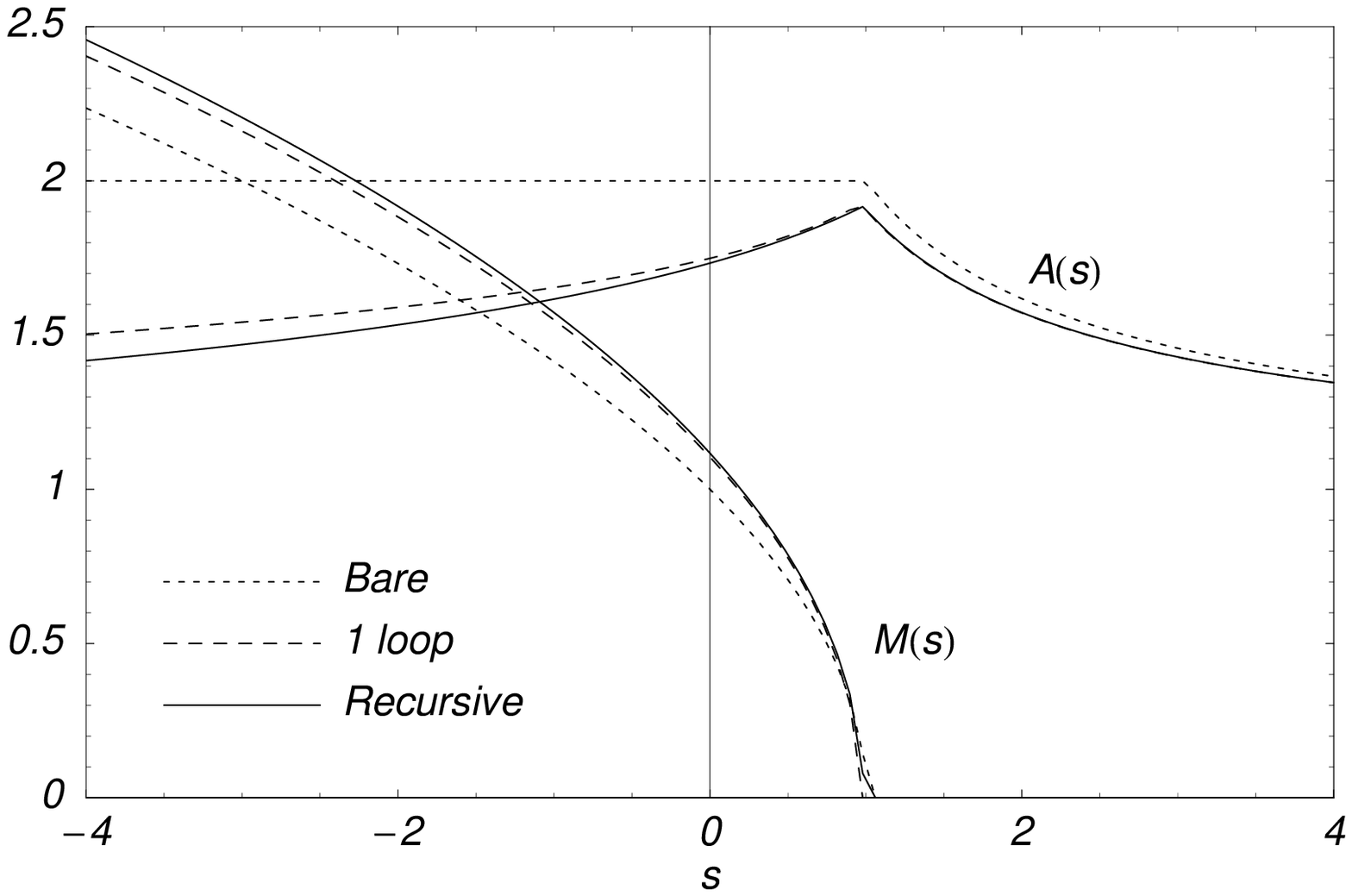}}

\centerline{\includegraphics[height=16em]{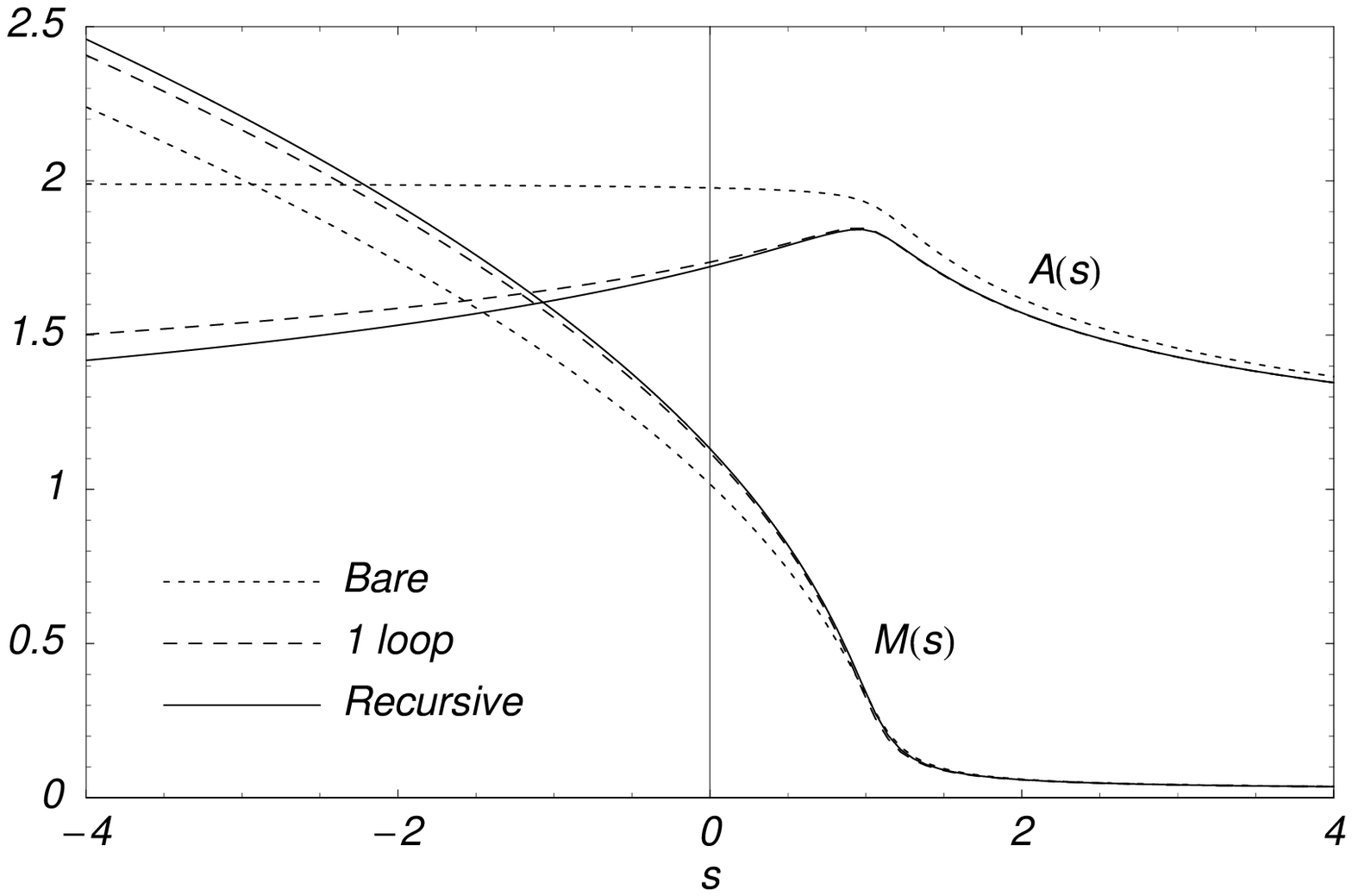}}

\caption{\label{AM0plot} Upper panel -- $A(s)$, $M(s)$ obtained with $m=0$:
solid line; Lower panel -- $A(s)$, $M(s)$ obtained with $m=0.023$: solid
line.  All dimensioned quantities are expressed in units of ${\cal G}$ in
Eq.~(\protect\ref{mnmodel}). A fit to meson observables requires ${\cal
G}\approx 0.5\,{\rm GeV}$ and hence this value of the current-quark mass
corresponds to $\approx 10\,$MeV.  In both panels, for comparison, we also
plot the results obtained with the zeroth-order vertex: dotted line; and the
one-loop vertex: dashed line.}
\end{figure} 
 
\begin{figure}[t] 

\centerline{\includegraphics[height=16em]{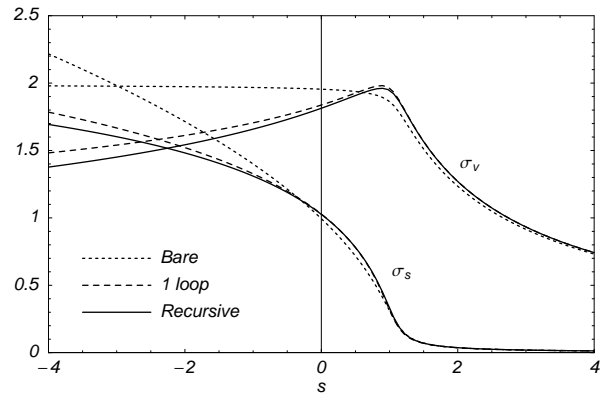}}

\caption{\label{SVSSmplot} The dressed-quark propagator can be written: 
$S(p)=-i\gamma\cdot p\,\sigma_V(s) + \sigma_S(s)$.  Here we plot $4\, 
\sigma_V(s)$, $2 \,\sigma_S(s)$ obtained with $m=0.023$ cf.\ the results 
obtained with the zeroth-order vertex: dotted line; and the one-loop vertex: 
dashed line.  Clearly, the analytic properties of the dressed-quark propagator 
are qualitatively unaffected by our dressing of the vertex.} 
\end{figure} 
 
\begin{figure}[t]

\centerline{\includegraphics[height=16em]{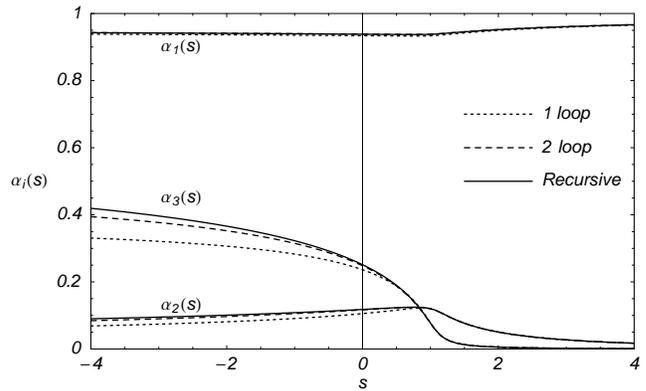}}

\caption{\label{alphamplot} $\alpha_i$, $i=1,2,3$, calculated with $m=0.023$.
These functions calculated at one-loop (dotted line) and two-loop (dashed
line) are also plotted for comparison.  The results obtained with $m=0$ are
little different.}
\end{figure} 
 
\begin{figure}[t] 

\centerline{\includegraphics[height=16em]{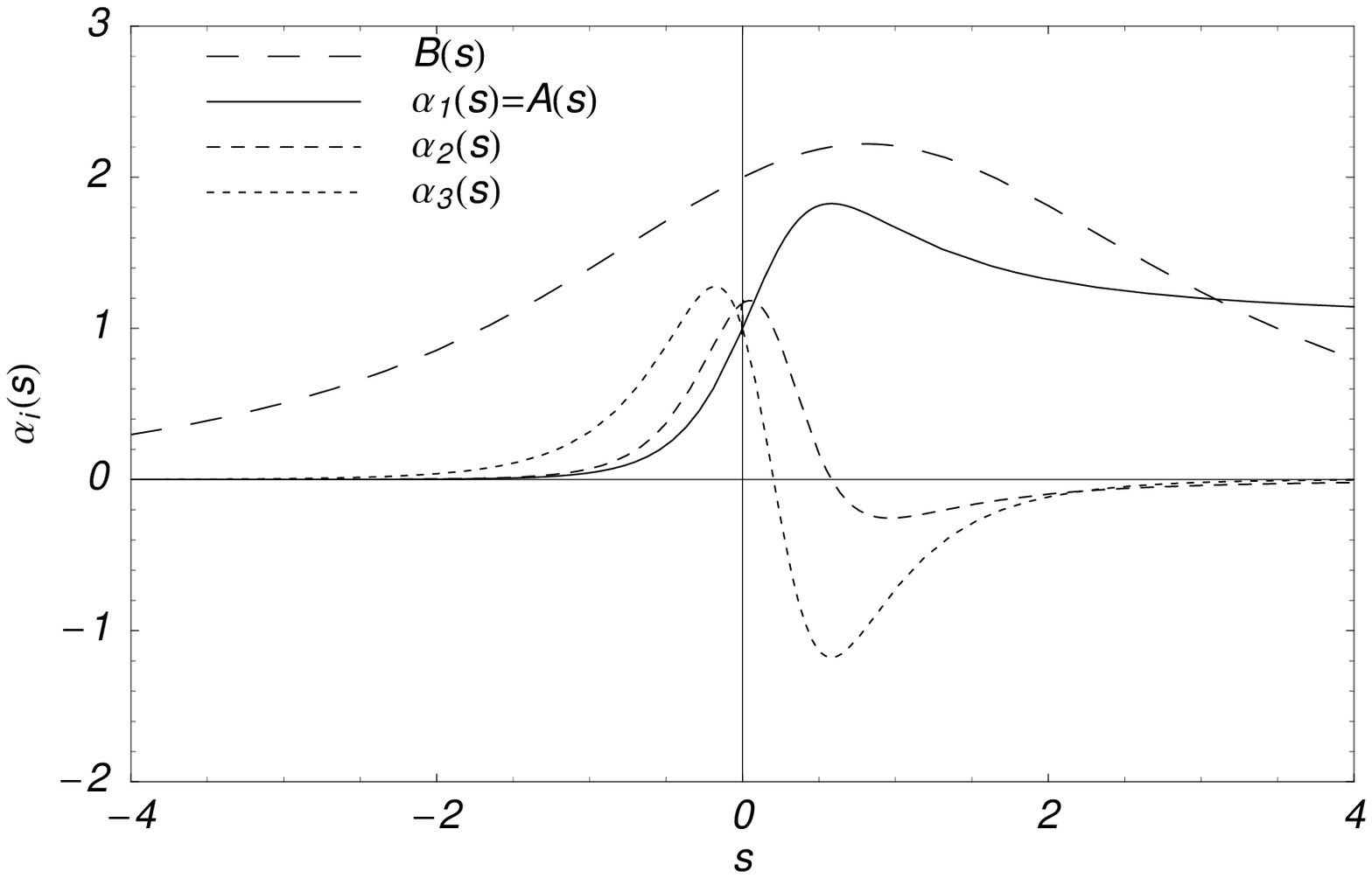}}

\caption{\label{bcplot} $\alpha_i$, $i=1,2,3$, calculated from 
Eqs.~(\protect\ref{bcvtx}) using the dressed-quark propagator solutions 
obtained in Ref.~\protect\cite{entireCJB}.} 
\end{figure} 
 
In Sec.~\ref{secthreeone} we described features of the Nambu-Goldstone mode
solution of the gap equation, whose $p^2\simeq 0$ properties are smoothly
related to those of the $m\neq 0$ solution, as is also the case in
QCD~\cite{mishaSVY}.  A complete solution is only available numerically and
our calculated results for the dressed-quark propagator and
gluon-ladder-dressed vertex are depicted in Figs.~\ref{AM0plot} --
\ref{alphamplot}.
 
It is apparent from the figures that the complete resummation of
dressed-gluon ladders yields a result for the dressed-quark propagator that
is little different from that obtained with the one-loop-corrected vertex;
and there is no material difference from the result obtained using the
zeroth-order vertex.  A single, exemplary quantification of this observation
is provided by a comparison between the values of $M(s=0)=B(0)/A(0)$
calculated using vertices dressed at different orders:
\begin{equation} 
\begin{array}{l|llll} 
%\displaystyle 
\sum_{i=0,N}\Gamma_\mu^i & N=0 & N=1 & N=2 & N=\infty\\[1.5ex]\hline 
M(0) & \rule{0em}{2ex} 1 & 1.105 &  1.115 & 1.117 
\end{array} 
\end{equation} 
 
Similar observations apply to the vertex itself.  Of course, there is a 
qualitative difference between the zeroth-order vertex and the 
one-loop-corrected result: $\alpha_{2,3}\neq 0$ in the latter case.  However, 
once that effect is seeded, the higher-loop corrections do little. 
 
\subsubsection{Vertex Ansatz} 
In the absence of a nonperturbatively dressed quark-gluon vertex a number of 
phenomenological DSE studies have employed an {\it Ansatz}, which is based on a 
nonperturbative analysis of QED and, in particular, is constrained by the 
vector Ward-Takahashi identity and the requirement of multiplicative 
renormalisability~\cite{BCMike}.  That vertex has been used~\cite{entireCJB} 
with the model interaction of Eq.~(\ref{mnmodel}) in which its form is 
expressed via 
\begin{equation} 
\label{bcvtx} \alpha_1(s) = A(s)\,,\; \alpha_2(s) = 2\,\frac{dA(s)}{ds}\,,\; 
\alpha_3(s) = 2\,\frac{dB(s)}{ds}\,. 
\end{equation} 
The functions defined by these expressions, calculated using the
self-consistent solutions determined in Ref.~\cite{entireCJB}, are depicted
in Fig.~\ref{bcplot}.  They bear little resemblance to the functions obtained
systematically via the resummation of dressed-gluon ladders, an outcome that
could be anticipated based on the difference between the dressed-quark
propagator calculated herein and that in Ref.~\cite{entireCJB}.  This finding
does not invalidate the {\it Ansatz} nor its use in modelling QCD but merely
shows that the {\it Ansatz}\ cannot be a sum of dressed-gluon ladders alone.
 
\section{Bethe-Salpeter Equation} 
\label{SecFour} 
\begin{figure}[t] 
 
\centerline{\includegraphics[height=7em]{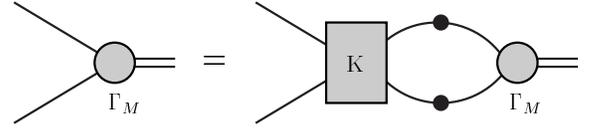}}
 
\caption{\label{BSEpic} Homogeneous BSE, Eq.~(\protect\ref{bsegen}).  Filled 
circles: dressed propagators or vertices; $K$ is the dressed-quark-antiquark 
scattering kernel.  A systematic truncation of $S^2 K$ is the key to preserving 
Ward-Takahashi identities~\protect\cite{truncscheme,herman}.} 
\end{figure}

The renormalised homogeneous Bethe-Salpeter equation (BSE) for the
quark-antiquark channel denoted by $M$ can compactly be expressed as
\begin{equation} 
\label{bsegen} 
    [\Gamma_M(k;P)]_{EF} =\int_q^\Lambda\! 
    [ K(k,q;P)]_{EF}^{GH}\, [\chi_M(q;P)]_{GH} 
\end{equation} 
where: $\Gamma_M(k;P)$ is the meson's Bethe-Salpeter amplitude, $k$ is the
relative momentum of the quark-antiquark pair and $P$ is their total
momentum; $E, \ldots ,H$ represent colour, flavour and spinor indices; and
\begin{equation} 
\chi_M(k;P) = S(k_+)\, \Gamma_M(k;P) \,S(k_-)\,. 
\end{equation} 
In Eq.~(\ref{bsegen}), which is depicted in Fig.~\ref{BSEpic}, $K$ is the
fully-amputated dressed-quark-antiquark scattering kernel.  The choice
\begin{equation} 
  [ K(k,q;P)]_{EF}^{GH} = {\cal D}_{\mu\nu}(k-q) \left[l^a 
\gamma_\mu\right]_{EG} \, \left[ l^a \gamma_\nu\right]_{HF}\,, 
\end{equation} 
yields the dressed-gluon ladder-truncation of the BSE, which provides the
foundation for many contemporary, field-theory-based phenomenological studies
of meson properties, see, e.g., Refs.~\cite{lucha}.
 
\subsection{Vertex-consistent Kernel} 
The preservation of Ward-Takahashi identities in those channels related to 
strong interaction observables requires a conspiracy between the 
dressed-quark-gluon vertex and the Bethe-Salpeter 
kernel~\cite{truncscheme,herman}.  We now describe a systematic procedure for 
building that kernel. 
 
As described, e.g., in Ref.~\cite{herman}, the DSE for the dressed-quark
propagator, $S$, is expressed via
\begin{equation} 
\frac{\delta \Gamma[S] }{\delta S} = 0 \,, 
\end{equation} 
where $\Gamma[S]$ is a Cornwall-Jackiw-Tomboulis-like effective action.  The 
Bethe-Salpeter kernel is then obtained via an additional functional derivative: 
\begin{equation} 
K_{EF}^{GH} = - \frac{\delta \Sigma_{EF}}{\delta S_{GH}}\,. \label{KCJT} 
\end{equation} 
 
Herein the self energy is given by the gap equation, Eq.~(\ref{gapmodel}), and 
the recursive nature of the dressed-gluon-ladder vertex entails that the $n$-th 
order contribution to the kernel is obtained from the $n$-loop contribution to 
the self energy: 
\begin{equation} 
    \Sigma^n(p)= - \int_q^\Lambda\! {\cal D}_{\mu\nu}(p-q) \, l^a \gamma_\mu \, 
    S(q) l^a \,\Gamma_{\nu}^n(q,p). 
\end{equation} 
Since $\Gamma_\mu(p,q)$ itself depends on $S$ then Eq.~(\ref{KCJT}) yields a 
sum of two terms:
\begin{eqnarray} 
\nonumber 
\lefteqn{ [K^n(k,q;P)]_{EF}^{GH} }\\ 
\nonumber & = & {\cal D}_{\mu\nu}(k - q) \left[ l^a \gamma_\mu \right]_{EG} 
\, 
\left[l^a\, \Gamma_\nu^n(q_-,k_-)\right]_{HF} \\ 
\nonumber & & +   \int_\ell^\Lambda\, {\cal D}_{\mu\nu}(k - \ell) \left[ 
l^a \gamma_\mu \,S(\ell_+) \right]_{EL}\, \\ 
& &  \;\;\;\;\;\;\;\; \times \frac{\delta }{\delta S_{GH}(q_\pm)} 
\left[l^a \,\Gamma_\nu^n(\ell_-,k_-)\right]_{LF}\,. 
\end{eqnarray} 
Here, in addition to the usual effect of differentiation, the functional 
derivative adds $P$ to the argument of every quark line through which it is 
commuted when applying the product rule.  
NB.\ ${\cal D}_{\mu\nu}$ also depends on $S$ because of quark vacuum
polarisation diagrams.  However, as noted in Ref.~\cite{truncscheme}, the
additional term arising from the derivative of ${\cal D}_{\mu\nu}$ does not
contribute to the BSE kernel for flavour non-diagonal systems, which are our
focus herein, and hence is neglected for simplicity.  It must be included
though to obtain a kernel adequate for an analysis of problems such as the
$\eta$-$\eta^\prime$ mass splitting, for example.

Now, introducing $[\chi_M(q;P)]_{GH}$, the BSE becomes
\begin{eqnarray} 
\nonumber \lefteqn{\Gamma_M(k;P)  =  \int_q^\Lambda 
{\cal D}_{\mu\nu}(k - q)\,}\\ 
\nonumber && \times \, l^a \gamma_\mu \left[\rule{0em}{3ex} \chi_M(q;P) \, 
l^a\, 
\Gamma_\nu(q_-,k_-) + S(q_+) \, \Lambda_{M \nu}^{a}(q,k;P)\right], \\ 
\label{genbsenL1} 
\end{eqnarray} 
where we have used Eq.~(\ref{vtxsummeda}) and defined 
\begin{equation} 
\Lambda_{M \nu}^{a}(q,k;P) = \sum_{n=0}^\infty \Lambda_{M \nu}^{a;n}(q,k;P)\,, 
\label{Lambdatotal} 
\end{equation} 
with 
\begin{eqnarray} 
\nonumber \lefteqn{ 
 \left[ \Lambda^{a;n}_{M \nu}(\ell,k;P) \right]_{LF} = }\\ 
\nonumber & & \int_q^\Lambda\!  \frac{\delta }{\delta S(q_\pm)_{GH}} 
\left[ l^a \,\Gamma_\nu^n(\ell_-,k_-)\right]_{LF} 
[\chi_M(q;P)]_{GH}\,.\\[-2ex] 
\label{defLambda}
\end{eqnarray} 
 
\begin{figure}[b] 

\centerline{\includegraphics[height=7em]{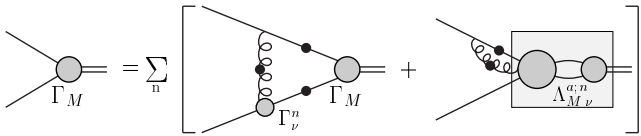}}

\caption{\label{BSE2} BSE expressed in Eq.~(\protect\ref{genbsenL1}), which is 
valid whenever $\Gamma_\mu$ can be obtained via a recursion relation.} 
\end{figure}

Equation~(\ref{genbsenL1}) is depicted in Fig.~\ref{BSE2}.  The first term is
instantly available once one has an explicit form for $\Gamma^{n}_{\nu}$.  To
develop an understanding of the second term, which is identified by the
shaded box in the figure, we employ the recursive expression for the
dressed-quark-gluon vertex, Eq.~(\protect\ref{vtxrecursfull}), with $p=k$,
$P=\ell - k$, to obtain an inhomogeneous recursion relation for
$\Lambda_{M \nu}^{a;n}$:
\begin{widetext} 
\begin{eqnarray} 
\nonumber \Lambda^{a;n}_{M \nu}(\ell,k;P) &= & \int_q^\Lambda\! {\cal 
D}_{\rho\sigma}(\ell-q)\,l^b \gamma_\rho\, \chi_M(q;P)\,  l^a 
\Gamma_\nu^{n-1}(q_-,q_-+k-\ell) 
\, S(q_-+k-\ell)\, l^b \gamma_\sigma\\ 
\nonumber && +\int_{q}^\Lambda\! {\cal D}_{\rho\sigma}(k -q)\, l^b
\gamma_\rho\,  
S(q_+ + \ell - k)\, l^a \Gamma_\nu^{n-1}(q_+ + \ell - k,q_+)\, \chi_M(q;P) 
\,l^b \gamma_\sigma \\
&& + 
\int_{q^\prime}^\Lambda\! {\cal D}_{\rho\sigma}(\ell -q^\prime) l^b 
\gamma_\rho\, S(q^\prime_+)\, \Lambda^{a;n-1}_{M \nu}(q^\prime,q^\prime 
+k - \ell;P)\, S(q_-^\prime+k-\ell) \, l^b \gamma_\sigma
\,. \label{Lambdarecursion}  
\end{eqnarray} 
\end{widetext} 
 
\begin{figure}[b] 
 
\centerline{\includegraphics[height=5.5em]{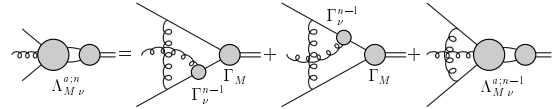}}

\caption{\label{picLambda} Recursion relation for $\Lambda_{M \nu}^{a;n}$,
Eq.~(\protect\ref{Lambdarecursion}).  We label the diagrams on the r.h.s.,
from left to right, as ${\cal G}_1$, ${\cal G}_2$, ${\cal L}$.}
\end{figure}

Equation~(\ref{Lambdarecursion}) is illustrated in Fig.~\ref{picLambda} and,
combined with Figs.~\ref{gamma_recursion}, \ref{BSE2}, this discloses the
content of the vertex-consistent Bethe-Salpeter kernel: namely, it consists
of a countable infinity of contributions, an infinite subclass of which are
crossed-ladder diagrams and hence nonplanar.  It is clear that every $n>0$,
vertex-consistent kernel must contain nonplanar diagrams.  Charge conjugation
can be used to expose a diagrammatic symmetry in the Bethe-Salpeter kernel,
which is the procedure we used, e.g., to obtain Eq.~(\ref{gapmodel}).  (The
steps and outcome described here formalise the procedure illustrated in
Fig.~1 of Ref.~\cite{truncscheme}.)
 
At this point if $\Gamma^{n}_{\nu}$ and the propagator functions: $A$, $B$,
are known, then $\Lambda_{M \nu}^{a}$, and hence the channel-projected
Bethe-Salpeter kernel, can be calculated explicitly.  That needs to be done
separately for each channel because, e.g., $\Lambda_{M \nu}^{a}$ depends on
$\chi_M(q;P)$.
 
To proceed we observe that the Bethe-Salpeter amplitude for a $\pi$-meson is 
\begin{eqnarray} 
\nonumber \lefteqn{ \Gamma_\pi^j(k;P) = \mbox{\boldmath $I$}_c\, 
\tau^j\,\gamma_5\,\left[ i 
f^1_\pi(k^2,k\cdot P;P^2) \right. }\\ 
\nonumber && +\, \gamma\cdot P\,f^2_\pi(k^2,k\cdot P;P^2) + \gamma\cdot k 
\,k\cdot 
P\, f^3_\pi(k^2,k\cdot P;P^2)\\ 
& & \left. + \,\sigma_{\mu\nu}k_\mu P_\nu\,f^4_\pi(k^2,k\cdot P;P^2) \right]\,, 
\label{pibsa} 
\end{eqnarray} 
where $\mbox{\boldmath $I$}_c$ is the identity in colour space (mesons are 
colour singlets) and $\{\tau^j;j=1,2,3\}$ are the Pauli matrices. This 
illustrates the general structure of meson Bethe-Salpeter amplitudes, which 
hereafter we express via 
\begin{equation} 
\label{genbsa} \Gamma_M(k;P) = \mbox{\boldmath $I$}_c \sum_{i=1}^{N_M}\, {\cal 
G}^i(k;P) \,f^i_M(k^2,k\cdot P;P^2)\,, 
\end{equation} 
where ${\cal G}^i(k;P)$ are the independent Dirac matrices required to span the 
space containing the meson under consideration.  (Subsequently, to simplify our 
analysis, we focus on $N_f=2$ and assume isospin symmetry. There is no 
impediment, in principle, to a generalisation.) 
 
The result of substituting Eq.~(\ref{genbsa}) into the r.h.s.\ of
Eq.~(\ref{genbsenL1}) can be expressed in the compact form
\begin{equation} 
\mbox{\boldmath $f$} = {\cal H}\mbox{\boldmath $f$} = ( {\cal H}_1 + {\cal
H}_2 )\mbox{\boldmath $f$} \,, \label{bse2}
\end{equation} 
where \mbox{\boldmath $f$} is a column vector composed of the scalar
functions: $\mbox{\boldmath $f$}_i=f^i_M$.  Here ${\cal H}_1$ is the
contribution from the first term on the r.h.s.\ in Eq.~(\ref{genbsenL1}) and
it is a $N_M \times N_M$ matrix wherein the elements of row-$j$ are obtained
via that Dirac trace projection which yields $f^j_M$ on the l.h.s.; i.e., for
${\cal P}_j$ such that
\begin{equation} 
f^j_M = {\rm tr}_D \left[ {\cal P}_j \Gamma_M \right] , \label{PGamma} 
\end{equation} 
then, using $l^a \,l^a = - C_2(R) = -\sfrac{4}{3}$ for $SU(N_c=3)$, 
\begin{eqnarray} 
\nonumber \lefteqn{ [{\cal H}_1]_{j,k} \mbox{\boldmath $f$}_k = 
-\sfrac{4}{3}\,{\rm tr}_D\left[ {\cal P}_j\, \int_\ell^\Lambda {\cal D} 
_{\mu\nu}(k - \ell) 
\gamma_\mu\, S(\ell_+)\,\right. }\\ 
\nonumber & & \times \, \left.\rule{0ex}{3ex} {\cal G}^k(\ell;P)\,S(\ell_-) \, 
\Gamma_\nu(\ell_-,k_-) \right]  f^k_M(\ell^2,\ell\cdot P;P^2)\,.\\ 
\end{eqnarray} 

${\cal H}_2$ represents the contribution from the second term, to which we
now turn.  In mesonic channels the colour structure of $\Lambda_{M
\nu}^{a;n}$ is simple:
\begin{equation} 
\label{colourfactor} \Lambda_{M\nu}^{a;n}  = l^a \, \Lambda_{M\nu}^{n} 
\end{equation} 
because $\chi_M \propto \mbox{\boldmath $I$}_c$ and $ l^b \, l^a \,l^b 
= l^a/(2 N_c)$.  The other part of the direct product is a matrix in Dirac 
space that can be decomposed as follows: 
\begin{eqnarray} 
\Lambda_{M\nu}^{n} & = & \sum_{\lambda=1}^{N_\Lambda} 
\beta_\lambda^{n}(\ell,k;P)\,g^\lambda_\nu(\ell,k;P) \label{Lambdaexpand}\\ 
&=& \sum_{\lambda=1}^{N_\Lambda}\,\sum_{j=1}^{N_M}\,f^j_M \, 
\beta_\lambda^{j;n}(\ell,k;P)\,g^\lambda_\nu(\ell,k;P)\,, 
\end{eqnarray} 
where the sum over $j$ implicitly expresses an integral over the relative
momentum appearing in $\chi_M$;
$\{g_\nu^\lambda;\lambda=1,\ldots,N_\Lambda\}$ are the independent Dirac
matrices required to completely describe $\Lambda_{M \nu}^{n}$, whose form
and number are determined by the structure of $\chi_M$, $N_\Lambda \geq N_M$;
and $\{\beta_\lambda^n(P^2); \lambda=1,\ldots,N_\Lambda\}$ are the associated
scalar coefficient functions.  (We subsequently suppress momentum arguments
and integrations for notational ease.)
 
Using Eq.~(\ref{Lambdaexpand}), the recursion relation of 
Eq.~(\ref{Lambdarecursion}) translates into a relation for 
$\{\beta_\lambda^n;i=\lambda,\ldots,N_\Lambda\}$.  To obtain that relation one 
first isolates these functions via trace projections. That can be achieved by 
using any complete set of projection operators: $\{{\cal 
P}_{\Lambda;\nu}^i;i=1,\ldots,N_\Lambda\}$, in which case one has 
\begin{equation} 
\beta_\lambda^n = [\mbox{\boldmath $M$}]_{\lambda \lambda^\prime}\, {\rm 
tr}_{CD} \left[\sfrac{1}{8i} \lambda^a \,{\cal 
P}_{\Lambda;\nu}^{\lambda^\prime}\,\Lambda_{M \nu}^{a;n}\right],
\label{traceLambda} 
\end{equation} 
where ${\rm tr}_{CD}$ identifies a trace over colour and Dirac indices.
(NB.\ The optimal choice of projection operators would yield
$[\mbox{\boldmath $M$}]_{\lambda \lambda^\prime} = \delta_{\lambda
\lambda^\prime}$, as we assumed, e.g., in Eq.~(\ref{PGamma}).) We
subsequently adopt a compact matrix representation of
Eq.~(\ref{traceLambda}):
\begin{equation} 
\mbox{\boldmath $\beta$}^n = \mbox{\boldmath $M$} \, \mbox{\boldmath ${\cal 
T}$}^n \,, \label{betarecursion} 
\end{equation} 
where $\mbox{\boldmath $\beta$}^n$ is a column vector with $N_\Lambda$ entries.
 
Replacing $\Lambda_{M \nu}^{a;n}$ on the r.h.s.\ in Eq.~(\ref{traceLambda})
by the r.h.s.\ of Eq.~(\ref{Lambdarecursion}) and using the distributive
property of the trace operation, one obtains
\begin{equation} 
\mbox{\boldmath ${\cal T}$}^n  = \mbox{\boldmath $G$}\,\mbox{\boldmath 
$\alpha$}^{n-1} + \mbox{\boldmath $L$}\, \mbox{\boldmath $\beta$}^{n-1}\,, 
\label{distributive} 
\end{equation} 
where \mbox{\boldmath $G$} describes the contributions to the trace from the
first two terms in Fig.~\ref{picLambda}, ${\cal G}_{1,2}$, which are
determined by the dressed-quark-gluon vertex and thus proportional to
$\mbox{\boldmath $\alpha$}^{n-1}$, and \mbox{\boldmath $L$} represents the
contribution from the last term, ${\cal L}$. Using Eq.~(\ref{distributive}),
Eq.~(\ref{betarecursion}) becomes
\begin{equation} 
\mbox{\boldmath $\beta$}^n = \mbox{\boldmath $M$}\,\left[\mbox{\boldmath 
$G$}\,\mbox{\boldmath $\alpha$}^{n-1} + \mbox{\boldmath $L$}\, \mbox{\boldmath 
$\beta$}^{n-1}\right]\,. \label{betaexpanded} 
\end{equation} 
(We reiterate that \mbox{\boldmath $M$}, \mbox{\boldmath $G$} and 
\mbox{\boldmath $L$} are all functionals of $\chi_M$ and hence are different 
for each meson channel.) 
 
It is evident that Eq.~(\ref{betaexpanded}) entails 
\begin{eqnarray} 
\nonumber \mbox{\boldmath $\beta$}^n & = &  [ \mbox{\boldmath 
$M$}\mbox{\boldmath $L$}]^n \,\mbox{\boldmath $\beta$}^0 + \sum_{j=0}^{n-1}\, [ 
\mbox{\boldmath $M$}\mbox{\boldmath $L$}]^j \, \mbox{\boldmath $M$} 
\mbox{\boldmath $G$}\,\,\mbox{\boldmath $\alpha$}^{n-j-1} \\ 
& = & \sum_{j=0}^{n-1}\, [ \mbox{\boldmath $M$}\mbox{\boldmath $L$}]^j \, 
\mbox{\boldmath $M$} \mbox{\boldmath $G$}\, {\cal O}^{n-j-1} \mbox{\boldmath 
$\alpha$}^{0}\,, 
\end{eqnarray} 
where ${\cal O}$ resolves the dressed-quark-gluon-vertex recursion, as we saw 
with our simple dressed-gluon model in connection with 
Eq.~(\ref{matrixrecurs}), and the first term vanishes because $\mbox{\boldmath 
$\beta$}^0=0$ by definition, Eq.~(\ref{defLambda}).
 
Finally, the complete BSE involves the sum expressed in
Eq.~(\ref{Lambdatotal}), which is determined by:
\begin{eqnarray} 
\nonumber \mbox{\boldmath $\beta$}  = \sum_{n=1}^\infty\, \mbox{\boldmath 
$\beta$}^n 
& = &\sum_{n=1}^\infty\,\sum_{j=0}^{n-1}\, [ \mbox{\boldmath 
$M$}\mbox{\boldmath $L$}]^j \, \mbox{\boldmath $M$} \mbox{\boldmath $G$}\, 
{\cal O}^{n-j-1} \mbox{\boldmath $\alpha$}^{0} \\ 
\nonumber & = & \sum_{i=0}^\infty\,[ \mbox{\boldmath $M$}\mbox{\boldmath 
$L$}]^i \, \mbox{\boldmath $M$} \mbox{\boldmath $G$}\, \sum_{j=0}^\infty\, 
{\cal O}^{j} \mbox{\boldmath $\alpha$}^{0} \\ 
& = & \frac{1}{1 - \mbox{\boldmath $M$}\mbox{\boldmath $L$}} \, \mbox{\boldmath 
$M$} \mbox{\boldmath $G$} \, \frac{1}{1 - {\cal O}} \, \mbox{\boldmath 
$\alpha$}^{0}\,, \label{fullbeta} 
\end{eqnarray} 
and this, via Eq~(\ref{Lambdaexpand}), completely determines the second term in 
Eq.~(\ref{genbsenL1}) so that we can complete Eq.~(\ref{bse2}) with 
\begin{eqnarray} 
\nonumber \lefteqn{ [{\cal H}_2]_{j,k} \mbox{\boldmath $f$}_k =
-\sfrac{4}{3}\,{\rm tr}_D\left[ {\cal P}_j\, \int_\ell^\Lambda {\cal
D}_{\mu\nu}(k - \ell) \right. }\\
& & \times \, \left.\rule{0ex}{3ex} \gamma_\mu\, S(\ell_+) \,
g_\nu^\lambda(\ell,k;P)\, \beta_\lambda^k(\ell,k;P) \right] f^k_M(P^2)\,,
\end{eqnarray} 
which we have now demonstrated is calculable in a closed form.  (Recall that
the sum over $k$ implicitly expresses an integral over the relative momentum
in the Bethe-Salpeter amplitude.  Note, too, that there is no sum over
iterates in this equation: it is \mbox{\boldmath $\beta$} from
Eq.~(\ref{fullbeta}) which appears.)
 
\subsection{Solutions of the vertex-consistent meson Bethe-Salpeter Equation} 
To elucidate the content of the BSE just derived we return to the algebraic 
model generated by Eq.~(\ref{mnmodel}).  In this case solutions of the BSE are 
required to have relative momentum $k=0$ so that Eq.~(\ref{genbsa}) simplifies 
to 
\begin{equation} 
\Gamma_M(P) = \sum_{i=1}^{N<N_M}\,{\cal G}^i(P)\,f_i(P^2)\, 
\end{equation} 
and Eq.~(\ref{bse2}) is truly algebraic; i.e., there are no implicit
integrations.  Furthermore, the kernel ${\cal H} = {\cal H}(P^2)$; i.e., it
is a matrix valued function of $P^2$ alone, and therefore the mass, $M_H^2$,
of any bound state solution is determined by the condition
\begin{equation} 
\label{characteristic} \left.\det\left[{\cal H}(P^2) - \mbox{\boldmath 
$I$}\right]\right|_{P^2+M_H^2=0} = 0\,, 
\end{equation} 
which is the requirement for any matrix equation: $H x = x$, to have a
nontrivial solution: it is the characteristic equation.  NB.\ If no solution
of Eq.~(\ref{characteristic}) exists then the model doesn't produce a bound
state in the channel under consideration.
 
\subsubsection{$\pi$-meson} 
\label{sssec:pion} 
In our algebraic model, because $k=0$, Eq.~(\ref{pibsa}) simplifies to
\begin{equation} 
\Gamma_\pi(P) = \gamma_5\,\left[ i f_1(P^2) + \gamma\cdot \hat
P\,f_2(P^2)\right],
\end{equation} 
where $\hat P$ is the direction-vector associated with $P$; $\hat P^2=1$, and
for the projection operators of Eq.~(\ref{PGamma}) we choose
\begin{equation} 
{\cal P}_1 = -\sfrac{i}{4} \,\gamma_5\,,\; 
{\cal P}_2 = \sfrac{1}{4}\, \gamma\cdot \hat P\,\gamma_5\,. 
\end{equation} 
 
The vertex $\Lambda_{\pi\nu}^{a;n}$ transforms as an axial-vector and its form 
is therefore spanned by twelve independent Dirac amplitudes.  However, since 
the relative momentum is required to vanish, that simplifies and we have: 
\begin{eqnarray} 
\nonumber \Lambda_{\pi\nu}^{a;n} & = & l^a\,\gamma_5 \left[\beta_1^n(P^2) 
\,\gamma_\nu + \beta_2^n(P^2)\,\gamma\cdot \hat P\,\hat P_\nu 
\right. \\ 
&& \left.  + \, \beta_3^n(P^2)\, \hat P_\nu + \beta_4^n(P^2)\, 
\gamma_\nu\,\gamma\cdot \hat P\right]\,, 
\end{eqnarray} 
and an obvious choice for the projection operators of Eq.~(\ref{traceLambda})
is
\begin{equation} 
\begin{array}{lclclcl} 
{\cal P}_{\Lambda_\pi; \nu}^1 &=& \sfrac{1}{16}\,\gamma_\nu\,\gamma_5\,, &\; 
{\cal P}_{\Lambda_\pi; \nu}^2 & = & \sfrac{1}{4}\,\hat P_\nu \, 
\gamma\cdot \hat P \, \gamma_5\,,\\ 
&&&&&& \\ 
{\cal P}_{\Lambda_\pi; \nu}^3 & = & \sfrac{1}{4}\, \hat P_\nu\, 
\gamma_5 \,, & \; 
{\cal P}_{\Lambda_\pi; \nu}^4 & = & \sfrac{1}{16}\, \gamma\cdot 
\hat P\, \gamma_\nu\, \gamma_5\,. 
\end{array} 
\end{equation} 
 
It can now be shown that $\mbox{\boldmath $G$}_\pi = 0$ in
Eq.~(\ref{distributive}): $\mbox{\boldmath $G$}_\pi$ is the sum of two terms,
the first and second in Eq.~(\ref{Lambdarecursion}), and using
Eq.~(\ref{mnmodel}) and recalling that this model forces the relative
momentum to vanish in the bound state amplitude, these two terms are equal in
magnitude but opposite in sign because the projection operators are
axial-vector in character.  (At leading order this result expresses an exact
cancellation between one of the one-loop vertex corrections and the
crossed-box term in Fig.~1 of Ref.~\cite{truncscheme}.) $\mbox{\boldmath
$G$}_\pi=0$ is a consequence of the fact that under charge conjugation
$\Lambda_{\pi \mu}(\ell,k;P)$ transforms according to:
\begin{equation} 
\bar\Lambda_{\pi \mu}(-k,-\ell;P)^{\rm t} = - \Lambda_{\pi \mu}(\ell,k;P)\,,
\end{equation} 
where $(\cdot)^{\rm t}$ denotes matrix transpose, coupled with the result that 
Eq.~(\ref{mnmodel}) enforces $k=0=\ell$ in the BSE.  It is not a general 
feature of the vertex-consistent Bethe-Salpeter kernel. 
 
It is plain from Eqs.~(\ref{Lambdaexpand}), (\ref{fullbeta}) that
$\mbox{\boldmath $G$}_\pi=0$ entails
\begin{equation} 
\label{Lambdais0} \Lambda_{\pi \nu}^a \equiv 0 
\end{equation} 
and hence the complete vertex-consistent pion BSE is simply $\mbox{\boldmath 
$f$} = {\cal H}_1\,\mbox{\boldmath $f$}$; i.e., 
\begin{equation} 
\Gamma_\pi(P) =- \,\gamma_\mu\, 
S(Q)\,\Gamma_\pi(P)\,S(-Q)\,\Gamma_\mu(-Q,-Q)\,, \label{pibsemnmodel} 
\end{equation} 
with $Q=P/2$, and the dressed-quark propagator and dressed-quark gluon vertex 
calculated in Sec.~\ref{secthreetwo}. 
 
\begin{figure}[t] 

\centerline{\includegraphics[height=16em]{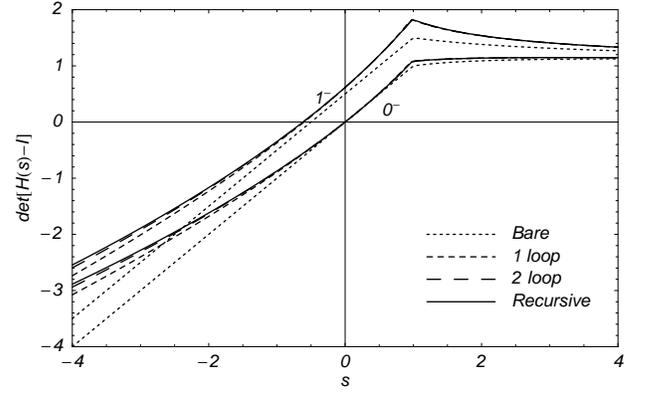}}

\caption{\label{pirho} Characteristic polynomial obtained in the chiral limit
from the vertex-consistent BSEs for the $\pi$- and $\rho$-mesons: the pion is
plainly massless.  The function obtained using $m=0.023$, which corresponds
to $\approx 10\,$MeV, is almost indistinguishable on the scale of this
figure. }
\end{figure}

The characteristic polynomial obtained from Eq.~(\ref{pibsemnmodel}) is
plotted in Fig.~\ref{pirho}: the zero gives the pion's mass,
Eq.~(\ref{characteristic}), which is listed in Table~\ref{masses}.
Figure~\ref{pirho} provides a forthright demonstration that the pion is
massless in the chiral limit.  This is a model-independent consequence of the
consistency between the Bethe-Salpeter kernel we have constructed and the
kernel in the gap equation. The figure illustrates that the vertex-consistent
Bethe-Salpeter kernel converges just as rapidly as the dressed-vertex itself
(cf.\ Fig.~\ref{alphamplot}). The insensitivity, evident in the table, of the
pion's mass to the order of the truncation is also explained by the fact that
our construction preserves the axial-vector Ward-Takahashi identity.
 
\begin{table}[b] 
\caption{\label{masses} Calculated $\pi$ and $\rho$ meson masses, in GeV,
quoted with ${\cal G}= 0.48\,{\rm GeV}$, in which case $m=0.023\, {\cal G} =
11\,$MeV.  (In the notation of Ref.~\protect\cite{truncscheme}, this value of
${\cal G}$ corresponds to $\eta = 0.96\,$GeV.) $n$ is the number of
dressed-gluon rungs retained in the planar vertex, see
Fig.~\protect\ref{Gamma_inf}, and hence the order of the vertex-consistent
Bethe-Salpeter kernel: the rapid convergence of the kernel is apparent in the
tabulated results.\vspace*{1ex}}
\begin{ruledtabular} 
\begin{tabular*} 
{\hsize} {l@{\extracolsep{0ptplus1fil}} 
|c@{\extracolsep{0ptplus1fil}}c@{\extracolsep{0ptplus1fil}} 
c@{\extracolsep{0ptplus1fil}}c@{\extracolsep{0ptplus1fil}}} 
%{l|c|c|c} 
% 
 & $M_H^{n=0}$ & $M_H^{n=1}$ & $M_H^{n=2}$ & $M_H^{n=\infty}$\\\hline 
$\pi$, $m=0$ & 0 & 0 & 0 & 0\\ 
$\pi$, $m=0.011$ & 0.152 & 0.152 & 0.152 & 0.152\\\hline 
$\rho$, $m=0$ & 0.678 & 0.745 & 0.754 & 0.754\\ 
$\rho$, $m=0.011$ & 0.695 & 0.762 & 0.770 & 0.770 
\end{tabular*} 
\end{ruledtabular} 
\end{table} 
 
\subsubsection{$\rho$-meson} 
The complete form of the Bethe-Salpeter amplitude for a vector meson in our 
algebraic model is 
\begin{equation} 
\Gamma_\rho^\lambda(P) = \gamma \cdot\epsilon^\lambda(P)\, f_1^\rho(P^2) + 
\sigma_{\mu\nu}\,\epsilon_\mu^\lambda(P)\, \hat P_\nu\, f_2^\rho(P^2) \,. 
\label{gammarho} 
\end{equation} 
This expression, which only has two independent functions, is much simpler
than that allowed by a more realistic interaction, wherein there are eight
terms.  Nevertheless, Eq.~(\ref{gammarho}) retains the amplitudes that, in
more sophisticated studies, are found to be dominant~\cite{pieterrho}. In
Eq.~(\ref{gammarho}), $\{\epsilon_\mu^\lambda(P);\, \lambda=-1,0,+1\} $ is
the polarisation four-vector:
\begin{equation} 
P\cdot \epsilon^\lambda(P)=0\,,\; \forall \lambda\,;\;\; 
\epsilon^\lambda(P)\cdot\epsilon^{\lambda^\prime}(P) = \delta^{\lambda 
\lambda^\prime}. 
\end{equation} 
Here the projection operators of Eq.~(\ref{PGamma}) are 
\begin{equation} 
{\cal P}_1^\lambda= \sfrac{1}{12} \,\gamma\cdot \epsilon^\lambda(P)\,,\; 
{\cal P}_2^\lambda= \sfrac{1}{12} \, 
\sigma_{\mu\nu}\,\epsilon_\mu^\lambda(P)\, \hat P_\nu \,. 
\end{equation} 
 
To solve the vertex-consistent BSE we need to calculate 
$\Lambda_{\rho\,\nu}^{a;n}(\epsilon^\lambda,P)$, which is defined in 
Eq.~(\ref{Lambdarecursion}) and depends on the Bethe-Salpeter amplitude in 
Eq.~(\ref{gammarho}).  We have from Eq.~(\ref{colourfactor}) that 
\begin{eqnarray} 
\Lambda_{\rho\,\nu}^{a;n}(\epsilon^\lambda,P) & = & l^a\, 
\Lambda_{\rho\,\nu}^{n}(\epsilon^\lambda,P)\,, \label{Lambdarhoa} 
\end{eqnarray} 
and in our algebraic model the Dirac structure is completely expressed through 
\begin{eqnarray} 
\nonumber \lefteqn{ \Lambda_{\rho\,\nu}^{n}(\epsilon^\lambda,P) =
\beta^n_{\rho 1}(P) \epsilon_\nu^\lambda + \beta^n_{\rho
2}(P)\,\epsilon_\nu^\lambda\,i\gamma\cdot\hat P}\\
\nonumber && +\, \beta^n_{\rho 3}(P)\, i\gamma\cdot \epsilon^\lambda\,\hat
P_\nu\, + \beta^n_{\rho
4}(P)\,\sigma_{\alpha\beta}\,\epsilon_\alpha^\lambda(P)\, \hat
P_\beta\,\gamma_\nu\\
&& + \,\beta^n_{\rho 5}(P)\,
i\sigma_{\alpha\beta}\,\epsilon_\alpha^\lambda(P)\, \hat P_\beta\,\hat P_\nu
+ \beta^n_{\rho 6}(P)\, \gamma\cdot \epsilon^\lambda\,\gamma_\nu\,,
\label{Lambdarho} 
\end{eqnarray} 
which has this simple form because
$\Lambda_{\rho\,\nu}^{n}(\epsilon^\lambda,P)$ cannot depend on the relative
momenta.  In this case an obvious choice for the projection operators of
Eq.~(\ref{traceLambda}) is
\begin{equation} 
\begin{array}{lcllcl} 
{\cal P}^1_{\Lambda_\rho;\nu} &= &\sfrac{1}{4}\,\epsilon_\nu^\lambda , & 
{\cal P}^2_{\Lambda_\rho;\nu} & = & - \sfrac{i}{4}\, \gamma\cdot \hat 
P\,\epsilon_\nu^\lambda , \\[1ex] 
{\cal P}^3_{\Lambda_\rho;\nu} &=& - \sfrac{i}{4} \, \hat P_\nu \,\gamma\cdot 
\epsilon^\lambda , & && \\[1ex] 
{\cal P}^4_{\Lambda_\rho;\nu} & =&  \sfrac{1}{4}\,\gamma_\nu\, 
\sigma_{\alpha\beta}\,\epsilon_\alpha^\lambda(P)\,\hat P_\beta\,, && \\[1ex] 
{\cal P}^5_{\Lambda_\rho;\nu} & = & - \sfrac{i}{4}\, \hat P_\nu\, 
\sigma_{\alpha\beta}\,\epsilon_\alpha^\lambda(P)\, \hat P_\beta\,, & 
{\cal P}^6_{\Lambda_\rho;\nu} & = & \sfrac{1}{4}\,\gamma_\nu\, \gamma\cdot 
\epsilon^\lambda\,. 
\end{array} 
\label{rhoproject} 
\end{equation} 
 
The substitution of Eqs.~(\ref{Lambdarho}), (\ref{rhoproject}) into
Eq.~(\ref{traceLambda}) yields
\begin{equation} 
\mbox{\boldmath $M$}_\rho = \frac{1}{2} \left[ \begin{array}{rrrrrr} 
3 & 0 & 0 & 0 & 1 & -1 \\ 
0 &  3 & -1& -1& 0& 0\\ 
0& -1& 3& 1& 0& 0\\ 
0& -1& 1& 1& 0& 0\\ 
1& 0& 0& 0& 3& -1\\ 
-1& 0& 0& 0& -1& 1 
\end{array} \right] 
\end{equation} 
which simply expresses the fact that, e.g., 
\begin{equation} 
\beta_{\rho 1}^n(P) = - \sfrac{1}{8} \,{\rm tr}_{CD} \left[l^a \left(
3\,{\cal P}_{\Lambda_\rho;\nu}^1 + {\cal P}_{\Lambda_\rho;\nu}^5 - {\cal
P}_{\Lambda_\rho;\nu}^6 \right)
\Lambda_\nu^{a;n}\right] \! . 
\end{equation} 

The next step is a determination of the matrix $\mbox{\boldmath $G$}_\rho$ in
Eq.~(\ref{distributive}), which gives the contribution to the kernel's
recursion relation from the first two terms in Fig.~\ref{picLambda}. That is
achieved by substituting Eq.~(\ref{Lambdarhoa}) into
Eq.~(\ref{Lambdarecursion}), and using Eq.~(\ref{mnmodel}) and its
consequences this yields
\begin{eqnarray} 
\nonumber \lefteqn{ 2\,\Delta(Q^2)\,\mbox{\boldmath $M$}_\rho\,\mbox{\boldmath 
$G$}_\rho = 
}\\ 
\nonumber && 
\left[ \begin{array}{ccc} 
2[ f_1^\rho(P) B(Q^2) + f_2^\rho\,Q A(Q^2)] & 0& 0\\ 
f_1^\rho(P)\,Q\,A(Q^2) -f_2^\rho(P)\,B(Q^2) & 0 & 0 \\ 
- f_1^\rho(P)\,Q\,A(Q^2) +f_2^\rho(P)\,B(Q^2) & 0 & 0\\ 
- f_1^\rho(P)\,Q\,A(Q^2) +f_2^\rho(P)\,B(Q^2) & 0 & 0\\ 
0 & 0 & 0 \\ 
0 & 0 & 0 \\ 
\end{array} \right]\\ 
&& 
\end{eqnarray} 
with $Q = \surd Q^2 = \sqrt{P^2}/2$, and $\Delta(s) = s A^2(s) + B^2(s)$ as
in Eq.~(\ref{alpharesults}).  It is immediately apparent that the Dirac
components associated with $\beta^n_{\rho 5}$, $\beta^n_{\rho 6}$ are
annihilated by this part of the interaction.  Furthermore, the complete
contribution to $\mbox{\boldmath $\beta$}^n_{\rho}$ is
\begin{equation} 
\mbox{\boldmath $M$}_\rho\,\mbox{\boldmath $G$}_\rho \times 
\left[ \begin{array}{c} \alpha_1^{n-1}(Q^2) \\ \alpha_2^{n-1}(Q^2) \\ 
\alpha_3^{n-1}(Q^2) \end{array} \right], 
\end{equation} 
and so it is evident that the subleading Dirac components of the
dressed-quark-gluon vertex do not contribute to the $\Lambda_\nu^{a}$-related
part of the vertex-consistent Bethe-Salpeter kernel: they are eliminated by
the last two columns in $\mbox{\boldmath $M$}_\rho \mbox{\boldmath
$G$}_\rho$.  (These two simplifications are a feature of our model.)
 
The final element we require is the matrix $\mbox{\boldmath $L$}_\rho$ in 
Eq.~(\ref{distributive}); i.e., the contribution to the kernel from the last 
term in Fig.~\ref{picLambda}.  That is obtained by substituting 
Eq.~(\ref{Lambdarhoa}) into the last term of Eq.~(\ref{Lambdarecursion}), 
which gives 
\begin{eqnarray} 
\nonumber \lefteqn{ 4\,\Delta^2\,\mbox{\boldmath $M$}_\rho\,\mbox{\boldmath 
$L$}_\rho = 
}\\ 
\nonumber & & 
\left[ \begin{array}{cccccc} 
2\,\Delta& 0 & 0 & 0 & 0 &  2\,\Delta\\ 
0& -\,\Delta & 0 & -2\,\Delta & 0 & 0 \\ 
0& 0 & 2 [Q^2 A^2 - B^2] & 2 B^2 & -2 Q A B 
& -2 Q A B \\ 
0& 0 & 0 & \Delta & 0 & 0 \\ 
0 & 0 & 0& 0 & 0 & 0\\ 
0 & 0 & 0& 0 & 0 & 0 \\ 
\end{array} \right]\\ 
&& 
\end{eqnarray} 
where the argument of each function is $Q^2$; e.g., $B= B(Q^2)$.  In our
representation $\mbox{\boldmath $M$}_\rho \mbox{\boldmath $L$}_\rho$ does not
exhibit an explicit dependence on $\Gamma_\rho(P)$.  That dependence is
acquired through the recursion relation since $\mbox{\boldmath
$\beta$}^1_{\rho} \propto \mbox{\boldmath $M$}_\rho \mbox{\boldmath
$G$}_\rho$, as is apparent from Eq.~(\ref{betaexpanded}).
 
Note that this part of the kernel also annihilates the Dirac components
associated with $\beta^n_{\rho 5}$, $\beta^n_{\rho 6}$ in
$\Lambda^a_{\rho;\nu}(\lambda,P)$.  Hence the active form of the vertex is
\begin{eqnarray} 
\nonumber \lefteqn{ \Lambda_{\rho\,\nu}^{n}(\epsilon^\lambda,P) = \beta^n_{\rho
1}(P) \epsilon_\nu^\lambda + \beta^n_{\rho
2}(P)\,\epsilon_\nu^\lambda\,i\gamma\cdot\hat P}\\
 && +\, \beta^n_{\rho 3}(P)\, i\gamma\cdot \epsilon^\lambda\,\hat P_\nu\, +
\beta^n_{\rho 4}(P)\,\sigma_{\alpha\beta}\,\epsilon_\alpha^\lambda(P)\, \hat
P_\beta\,\gamma_\nu , \label{Lambdarhotrue}
\end{eqnarray} 
where the $\beta^n_{i}$, $i=1,\ldots,4$ are obtained from Eq.~(\ref{fullbeta}) 
using the elements calculated above. 
 
Putting all this together we have the complete vertex-consistent BSE for the
$\rho$-meson:
\begin{eqnarray} 
\nonumber \Gamma_\rho^\lambda(P) & = & - \,\gamma_\mu\, 
S(Q)\,\Gamma_\rho^\lambda(P)\,S(-Q)\,\Gamma_\mu(-Q,-Q) \\ 
& & - \, \gamma_\mu \, S(Q)\, \Lambda_{\rho\,\mu}(\epsilon^\lambda;P)\,, 
\label{rhoBSEfull} 
\end{eqnarray} 
with the dressed-quark propagator and dressed-quark gluon vertex calculated
in Sec.~\ref{secthreetwo}, and $\Lambda_{\rho \mu}(0,0;P)$ obtained from
Eqs.~(\ref{Lambdatotal}), (\ref{Lambdaexpand}), (\ref{fullbeta}) using the
matrices displayed above.
 
The characteristic polynomial obtained from Eq.~(\ref{rhoBSEfull}) is plotted
in Fig.~\ref{pirho}.  Its zero gives the $\rho$-meson mass and that is listed
in Table~\ref{masses}.  The tabulated values demonstrate that the
rainbow-ladder truncation underestimates the result of the complete
calculation by $\lesssim 10$\% (i.e., the calculation performed using the
fully-resummed dressed-gluon-ladder vertex) and simply including the
consistent one-loop corrections to the quark-gluon-vertex and Bethe-Salpeter
kernel reduces that discrepancy to $\lesssim 1$\%.  Furthermore, it is clear
that at every order of truncation the bulk of $m_\rho$ is obtained in the
chiral limit, which emphasises that the $\pi$-$\rho$ mass-splitting is driven
by the DCSB mechanism.
 
\subsubsection{Scalar and Axial-vector Mesons} 
In a simple constituent-quark picture, the ground state scalar and axial
vector mesons are angular-momentum $L=1$ eigen states.  This qualitative
feature is expressed in their Poincar\'e covariant Bethe-Salpeter amplitudes
through the presence of materially important relative-momentum-dependent
Dirac components; e.g., Refs.~\cite{angularmomentum}.  However, the model
defined by Eq.~(\ref{mnmodel}) forces meson Bethe-Salpeter amplitudes to be
independent of the constituent's relative momentum and, owing primarily to
that, the rainbow-ladder truncation of the model generates neither scalar nor
axial-vector meson bound states~\cite{mn83}.  We have found, unsurprisingly,
that improving the description of the dressed-quark-gluon vertex and
Bethe-Salpeter kernel is insufficient to overcome this defect of the model.
However, we anticipate that these improvements will materially alter the
results of BSE studies of scalar mesons that employ more realistic
interactions; e.g., Refs.~\cite{scalars}.
 
\subsection{Diquark Bethe-Salpeter Equation} 
The ladder-rainbow truncation, which is obtained by keeping only the first
term in Eq.~(\ref{vtxexpand}), generates colour-antitriplet quark-quark
(diquark) bound states~\cite{regdqmass}.  Such states are not observed in the
hadron spectrum, and it was demonstrated in Ref.~\cite{truncscheme} that they
are not present when one employs the one-loop-dressed vertex and consistent
Bethe-Salpeter kernel.  Herein we can verify that this feature persists with
the complete planar vertex and consistent kernel.  To continue, however, we
must slightly modify the procedure described above because of the
colour-antitriplet nature of the diquark correlation.  (NB.\ Colour-sextet
states are not bound in any truncation because even single gluon exchange is
repulsive in this channel. It is for this reason, too, that colour-octet
mesons do not appear.  Note also that the absence of colour-antitriplet
diquark bound states does not preclude the possibility that correlations in
this channel may play an important role in nucleon structure~\cite{regfe}
since some attraction does exist, e.g., Ref.~\cite{latticedq}.)
 
The analogue of Eq.~(\ref{bsegen}) is 
\begin{equation} \label{bsegendq} 
    [\Gamma_D(k;P)]_{EF} =\int_q^\Lambda\! 
    [ \bar K(k,q;P)]_{EF}^{GH}\, [\chi_D(q;P)]_{GH} 
\end{equation} 
where: $k$ is the relative momentum of the quark-quark pair and $P$ is their 
total momentum, as before; and 
\begin{equation} 
\chi_D(q;P) = S(q_+)\, \Gamma_D(q;P) \,S^{\rm t}(-q_-) 
\end{equation} 
with $\Gamma_D(q;P)$ the putative diquark's Bethe-Salpeter amplitude.  In this 
case $\bar K$ is the fully-amputated dressed-quark-quark-scattering kernel, for 
which 
\begin{equation} 
\bar K(k,q;P)]_{EF}^{GH} = {\cal D}_{\mu\nu}(k-q) \left[l^a 
\gamma_\mu\right]_{EG} \, \left[ (l^a \gamma_\nu)^{\rm 
t}\right]_{HF} 
\label{diquarkLadder} 
\end{equation} 
yields the dressed-gluon ladder-truncation of the BSE. 
 
Inspection of the general structure of $\bar K$ reveals that, following our 
ordering of diagrams, it can be obtained directly from the kernel in the meson 
BSE via the replacement 
\begin{equation} 
\label{diquarkRL} S(k)\,l^a\,\gamma_\mu \to \left[ \gamma_\mu\,l^a\,S(-k) 
\right]^{\rm t} 
\end{equation} 
in each antiquark segment of $K$, which can be traced unambiguously from the 
external antiquark line of the meson's Bethe-Salpeter amplitude. 
 
The appearance here of a matrix transpose makes the definition of a modified
Bethe-Salpeter amplitude~\cite{regdqmass} useful:
\begin{equation} 
\label{GammadqC} \Gamma_D^C(k;P) := \Gamma_D(k;P) \, C^\dagger\,, 
\end{equation} 
where $C= \gamma_2\gamma_4$ is the charge conjugation matrix, from which it 
follows that 
\begin{eqnarray} 
 \chi_D^C(k;P)&=& S(k_+)\,\Gamma_D^C(k;P)\, S(k_-)\,, 
\end{eqnarray} 
using $C \gamma_\mu^{\rm t} C^\dagger = -\gamma_\mu$.  $\Gamma_D^C(k;P)$ 
satisfies a BSE whose Dirac structure is identical to that of the meson BSE. 
However, its colour structure is different, with a factor of $(-l^a)^{\rm 
t}$ replacing $l^a$ at every gluon vertex on what was the conjugate-quark 
leg.  (NB.\ It is from this modification that the absence of diquark bound 
states must arise using the dressed-ladder vertex, if it arises at all.) For 
example, if one has a term in the meson BSE of the form 
\begin{equation} 
l^a \gamma_\mu \, S \, l^b \gamma_\nu \, S\,l^c \gamma_\rho \, S\, \Gamma_M S 
\, l^a \gamma_\mu \, S \, l^c \gamma_\rho \, S\,l^b \gamma_\nu \, 
\end{equation} 
then the related term in the equation for $\Gamma_D^C(k;P)$ is
\begin{equation} 
l^a \gamma_\mu \, S \, l^b \gamma_\nu \, S\,l^c \gamma_\rho \, S\, \Gamma_D^C S 
\, (-l^a)^{\rm t} \gamma_\mu \, S \, (-l^c)^{\rm t} \gamma_\rho \, 
S\,(-l^b)^{\rm t} \gamma_\nu \,. 
\end{equation} 
 
There are three colour-antitriplet diquarks and their colour structure is 
described by the matrices 
\begin{equation} 
\{\lambda^k_\wedge\,; k=1,2,3: \lambda^1_\wedge=\lambda^7\!,\, 
\lambda^2_\wedge=\lambda^5\!,\, \lambda^3_\wedge=\lambda^2 \}\,, 
\end{equation} 
and, as for mesons, their Bethe-Salpeter amplitudes can be written as a direct 
product, colour~$\otimes$~Dirac: 
\begin{equation} 
\Gamma_D^C(k;P)= \lambda_\wedge^k\, \Gamma_{qq}^C(k;P)\,. 
\end{equation} 
For colour-singlet mesons the colour factor is simply the identity matrix 
$\mbox{\boldmath $I$}_c$.  (Recall that we focus on $N_f=2$ and assume isospin 
symmetry.  Hence the diquark's flavour structure, which is described by the 
Pauli matrix $\tau^2$ in this case, cancels in the BSE.) 
 
As we observed, the rainbow-ladder diquark BSE is obtained by using 
Eq.~(\ref{diquarkLadder}) in Eq.~(\ref{bsegendq}).  Right-multiplying the 
equation thus obtained by $C^\dagger$ we find immediately that the equation 
satisfied by $\Gamma_D^C(k;P)$ is the same as the rainbow-ladder meson BSE 
\textit{except} that 
\begin{eqnarray} 
l^a \, \mbox{\boldmath $I$}_c\, l^a = - \,\sfrac{4}{3} \,\mbox{\boldmath 
$I$}_c & \rightarrow & l^a \,\lambda_\wedge^k\,(-l^a)^{\rm t} = 
-\,\sfrac{2}{3}\,\lambda_\wedge^k\,. 
\end{eqnarray} 
In both cases the colour matrix now factorises and can therefore be
cancelled.  Hence the rainbow-ladder BSEs satisfied by the colour-independent
parts of $\Gamma_M$ and $\Gamma_D^C(k;P)$ are identical but for a $50$\%
reduction of the coupling in the diquark equation. This expresses the fact
that ladder-like dressed-gluon exchange between two quarks is attractive and
explains the existence of diquark bound states in this
truncation~\cite{regdqmass}.
 
\begin{widetext} 
Following the above discussion it is apparent that the diquark counterpart of
Eq.~(\ref{Lambdarecursion}) is
\begin{eqnarray} 
\nonumber \Lambda^{a;n}_{D\nu}(\ell,k;P) &= & \int_q^\Lambda\! {\cal 
D}_{\rho\sigma}(\ell-q)\,l^b \gamma_\rho\, \chi_D^C(q;P)\, 
\Gamma_\nu^{n-1}(q_-,q_-+k-\ell)\,(l^a)^{\rm t} 
\, S(q_-+k-\ell)\, (l^b)^{\rm t} \gamma_\sigma\\ 
\nonumber & &- \int_{q}^\Lambda\! {\cal D}_{\rho\sigma}(k -q)\, l^b
\gamma_\rho\, S(q_+ + \ell - k)\, l^a \Gamma_\nu^{n-1}(q_+ + \ell - k,q_+)\,
\chi_D^C(q;P) \,(l^b)^{\rm t} \gamma_\sigma \\ 
& &- 
\int_{q^\prime}^\Lambda\! {\cal D}_{\rho\sigma}(\ell -q^\prime) l^b 
\gamma_\rho\, S(q^\prime_+)\,  \Lambda^{a;n-1}_{D\nu}(q^\prime,q^\prime 
+k - \ell;P)\, S(q_-^\prime+k-\ell) \, (l^b)^{\rm t} \gamma_\sigma\,.
\label{DLambdarecursion} 
\end{eqnarray} 
The factorisation of the colour structure observed in the ladder-like diquark
BSE persists at higher orders and can be used to obtain a recursion relation
for the diquark kernel analogous to that depicted in Fig.~\ref{picLambda}.
Indeed, a consideration of Eq.~(\ref{DLambdarecursion}) reveals that in
general one can write
\begin{equation} 
\Lambda^{a;n}_{D\nu}(\ell,k;P) = 
\Lambda^{n}_{1\nu}(\ell,k;P)\,l^a\,\lambda_\wedge^k + 
\Lambda^{n}_{2\nu}(\ell,k;P)\,\lambda_\wedge^k \,(l^a)^{\rm t}\,, 
\end{equation} 
where in this direct product $\Lambda^{n}_{1\nu}$ and $\Lambda^{n}_{2\nu}$ are 
Dirac matrices.  Defining: 
\begin{eqnarray} \nonumber \lefteqn{ {\cal T}^{L;n}_\nu 
\,\lambda_\wedge^k := l^a\,\Lambda^{a;n}_{D\nu} = - \, \sfrac{2}{3} \left[2 
\Lambda_{1\nu}^n - 
\Lambda_{2\nu}^n \right] \lambda_\wedge^k} \\ 
\nonumber & = & \left\{ - \,\sfrac{5}{9}\,\int_q^\Lambda\! {\cal 
D}_{\rho\sigma}(\ell-q)\,\gamma_\rho\, \chi_{qq}^C(q;P)\, 
\Gamma_\nu^{n-1}(q_-,q_-+k-\ell)\, S(q_-+k-\ell)\, 
\gamma_\sigma \right.\\ 
 && \nonumber - \sfrac{1}{9}\,\int_{q}^\Lambda\! {\cal D}_{\rho\sigma}(k 
-q)\, \gamma_\rho\, S(q_+ + \ell - k)\,  \Gamma_\nu^{n-1}(q_+ + \ell - k,q_+)\, 
\chi_{qq}^C(q;P) \, \gamma_\sigma  \\ 
& & \left. - 
\int_{q^\prime}^\Lambda\! {\cal D}_{\rho\sigma}(\ell -q^\prime) \gamma_\rho\, 
S(q^\prime_+)\, \left[\sfrac{1}{9} \, \Lambda^{n-1}_{1\nu}(q^\prime,q^\prime +k 
- \ell;P) - \sfrac{5}{9}\,  \Lambda^{n-1}_{2\nu}(q^\prime,q^\prime +k - 
\ell;P)\right] 
S(q_-^\prime+k-\ell) \, \gamma_\sigma\rule{0em}{3ex}\right\} 
\lambda_\wedge^k , \label{TLrecursion}\\ 
\nonumber \lefteqn{{\cal T}^{R;n}_\nu \,\lambda_\wedge^k := 
\Lambda^{a;n}_{D\nu} \, (l^a)^{\rm t} = - \, \sfrac{2}{3} \left[- 
\Lambda_{1\nu}^n + 2 \,\Lambda_{2\nu}^n \right] \lambda_\wedge^k }\\ 
\nonumber & = & \left\{\sfrac{1}{9}\,\int_q^\Lambda\! {\cal 
D}_{\rho\sigma}(\ell-q)\,\gamma_\rho\, \chi_{qq}^C(q;P)\, 
\Gamma_\nu^{n-1}(q_-,q_-+k-\ell)\, S(q_-+k-\ell)\, 
\gamma_\sigma \right. \\ 
& & + \sfrac{5}{9}\,\int_{q}^\Lambda\! {\cal D}_{\rho\sigma}(k 
-q)\, \gamma_\rho\, S(q_+ + \ell - k)\, \Gamma_\nu^{n-1}(q_+ + \ell - k,q_+)\, 
\chi_{qq}^C(q;P) \, \gamma_\sigma  \nonumber \\ 
& & \left. + \int_{q^\prime}^\Lambda\! {\cal D}_{\rho\sigma}(\ell
-q^\prime) \gamma_\rho\, S(q^\prime_+)\,\left[\sfrac{5}{9} \,
\Lambda^{n-1}_{1\nu}(q^\prime,q^\prime +k - \ell;P) -\sfrac{1}{9}\,
\Lambda^{n-1}_{2\nu}(q^\prime,q^\prime +k - \ell;P)\right]
S(q_-^\prime+k-\ell) \, \gamma_\sigma\rule{0em}{3ex}\right\}
\lambda_\wedge^k ,
\label{TRrecursion} 
\end{eqnarray} 
\end{widetext} 
then 
\begin{equation} 
\Lambda_{1\nu}^n = -\,\sfrac{1}{2}\,[ {\cal T}^{L;n}_\nu + 2\,{\cal 
T}^{R;n}_\nu ]\,,\; 
\Lambda_{2\nu}^n = -\,\sfrac{1}{2}\,[ 2 {\cal T}^{L;n}_\nu + {\cal T}^{R;n}_\nu 
]\,. 
\end{equation} 
NB.\ The first lines in each of Eqs.~(\ref{TLrecursion}), (\ref{TRrecursion})
prove that there is no mixing between colour-antitriplet and colour-sextet
diquarks, whose colour structure is described by the six symmetric Gell-Mann
matrices.
 
We can now write the vertex-consistent BSE for the colour-antitriplet diquark
channels (cf.\ Eq.~(\ref{genbsenL1})):
\begin{eqnarray} 
\nonumber \lefteqn{\Gamma_{qq}^C(k;P)  = - \sfrac{2}{3}\,\sum_{n=0}^{\infty}\, 
\int_\ell^\Lambda \! {\cal D}_{\mu\nu}(k - \ell)\,  }\\ 
\nonumber 
 && \times \gamma_\mu\left[ \chi_{qq}^C(\ell;P) \,\Gamma_\nu^n(\ell_-,k_-) - 
S(\ell_+)\,{\cal T}^{L;n}_\nu(\ell,k;P)  \right].\\ 
\label{genbseD} 
\end{eqnarray} 
It is straightforward to verify that this equation reproduces the diagrams 
considered explicitly in Ref.~\cite{truncscheme}. 
 
Having factorised the colour structure, the summation appearing in the first
term of Eq.~(\ref{genbseD}) yields the dressed vertex we have already
calculated.  One proceeds with the second term by analogy with
Eq.~(\ref{Lambdaexpand}) and observes that the matrix-valued functions,
$\Lambda_{1,2}$, can be decomposed:
\begin{equation} 
\Lambda_{i\,\nu}^n = 
\sum_{\lambda=1}^{N_{\Lambda_D}}\,\eta_{i\lambda}^n(\ell,k;P)\, 
h_\nu^\lambda(\ell,k;P)\,, 
\end{equation} 
where $\{h_\nu^\lambda\,;\lambda=1,\ldots,N_{\Lambda_D}\}$ is the smallest
set of Dirac matrices capable of expressing $\Lambda_{D\nu}^{a;n}$
completely, whose form and number depend on the channel under
consideration. Projection operators, ${\cal P}^{A}_{D;\nu}$, are easily
constructed so that, with
\begin{equation} 
\mbox{\boldmath $\eta$} := {\rm 
column}(\eta_{11},\,\ldots,\eta_{1N_{\Lambda_D}}, 
\eta_{21},\ldots,\eta_{2N_{\Lambda_D}})\,, 
\end{equation} 
we have 
\begin{equation} 
[\mbox{\boldmath $\eta$}^n]_A = [\mbox{\boldmath $M$}^\prime]_{AA^\prime}
{\rm tr}_D\left[{\cal P}^{A^\prime}_{D;\nu}\,\Lambda_{I\nu}^n\right]\,,
\end{equation} 
where $A,A^\prime = 1,\ldots,2N_{\Lambda_D}$, and $I=1$ for $A^\prime \leq 
N_{\Lambda_D}$ and $I=2$ for $A^\prime > N_{\Lambda_D}$.  Now the procedure of 
Eqs.~(\ref{betarecursion}) - (\ref{betaexpanded}) can be repeated to arrive at 
\begin{equation} 
\mbox{\boldmath $\eta$}^n =\mbox{\boldmath $M$}^\prime\,\left[ \mbox{\boldmath 
$G$}^\prime\, \mbox{\boldmath $\alpha$}^{n-1} + \mbox{\boldmath $L$}^\prime \, 
\mbox{\boldmath $\eta$}^{n-1} \right]\,, 
\end{equation} 
where $\mbox{\boldmath $G$}^\prime$, $\mbox{\boldmath $L$}^\prime$ are natural 
analogues of the matrices introduced in Eq.~(\ref{distributive}), and 
thereafter one can continue to obtain an obvious extension of 
Eq.~(\ref{fullbeta}). This completely determines the second term in 
Eq.~(\ref{genbseD}) and thus we have arrived at the vertex-consistent BSE for 
the colour-antitriplet diquark channel. 
 
\subsection{Solutions of the Diquark Equation} 
 
\subsubsection{Scalar Diquark} 
\begin{figure}[t] 

\centerline{\includegraphics[height=16em]{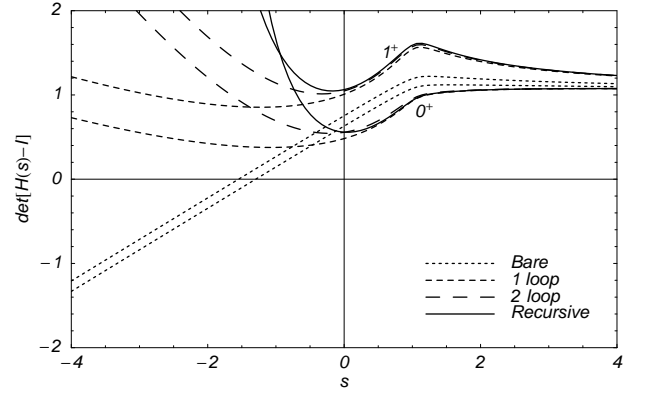}}

\caption{\label{Figdiquark} The characteristic polynomial obtained using
$m=0.023$, which corresponds to $\approx 10\,$MeV, calculated for the scalar
and axial-vector colour-antitriplet diquark channels using the diquark BSE,
Eq.~(\protect\ref{genbseD}).}
\end{figure}

In the algebraic model specified by Eq.~(\ref{mnmodel}) the general Dirac 
structure of a $J^P=0^+$ quark-quark correlation is 
\begin{equation} 
\Gamma^{0^+}_{qq}(P) = \gamma_5\,\left[ f_1^{0^+}(P^2) + \gamma\cdot P 
f_2^{0^+}(P^2)\right]\,, 
\end{equation} 
which is the same as that of the pion, for reasons which are obvious given
the discussion following Eq.~(\ref{GammadqC}).  The characteristic equation
for this channel is obtained following the method made explicit in
Sec.~\ref{sssec:pion} and it is depicted in Fig.~\ref{Figdiquark}.  The
existence of a bound state for $n=0$; i.e., in the rainbow-ladder truncation,
is apparent.  However, so too is the effect of the higher-order terms, which
was identified in Ref.~\cite{truncscheme}: at each higher-order nonplanar
diagrams in the kernel provide significant repulsion, which overwhelms any
attraction at that and preceding orders and thereby ensures diquark
confinement; i.e., the absence of coloured quark-quark bound states in the
spectrum. This feature is retained by the completely resummed kernel, using
which, instead of a zero, the characteristic polynomial exhibits a pole: the
repulsion is consummated. (This feature is not tied to the interaction in
Eq.~(\ref{mnmodel}); e.g., Ref.~\cite{dqcnfnjl}.)
 
\subsubsection{Axial Vector Diquark} 
The general form of a $J^P=1^+$ quark-quark correlation is represented by
\begin{equation} 
\Gamma_{qq}^{1^+}(P) = \gamma \cdot\epsilon^\lambda(P)\, f_1^{1^+}(P^2) + 
\sigma_{\mu\nu}\,\epsilon_\mu^\lambda(P)\, \hat P_\nu\, f_2^{1^+}(P^2) \,. 
\label{gammaavdq} 
\end{equation} 
Calculating the characteristic polynomial for this channel is also a
straightforward application of methods already introduced and it is plotted
in Fig.~\ref{Figdiquark}.  The features in this channel are qualitatively
identical to those of the scalar diquark.
 
\section{Summary} 
\label{SecFive} 
Using a planar quark-gluon vertex obtained through the resummation of
dressed-gluon ladders we have explicitly demonstrated that from a
dressed-quark-gluon vertex, obtained via an enumerable series of terms, it is
always possible to construct a vertex-consistent Bethe-Salpeter kernel that
ensures the preservation of Ward-Takahashi identities in the physical
channels related to strong interaction observables.  While we employed a
rudimentary model to make the construction transparent, the procedure is
general. However, the algebraic simplicity of the analysis is peculiar to our
model. For example, using a more realistic interaction the gap and vertex
equations would yield a system of twelve coupled integral equations.
Nevertheless, we anticipate that the qualitative features highlighted herein
are robust.
 
The simple interaction we employed characterises a class of models in which
the kernel of the gap equation has sufficient integrated strength to support
dynamical chiral symmetry breaking (DCSB).  The complete ladder summation of
this interaction, calculated self-consistently with the solution of the gap
equation, produces a dressed-vertex that is little changed cf.\ the bare
vertex. In particular, it does not exhibit an enhancement in the vicinity of
$k^2=0$, where $k$ is the momentum carried by the model dressed-gluon.  In
addition, the dressed-quark propagator obtained in this self-consistent
solution is qualitatively indistinguishable from that obtained using the
rainbow truncation.
 
The vertex-consistent Bethe-Salpeter kernel is necessarily nonplanar, even
when the vertex itself is planar, and in our simple model it is easily
calculable in a closed form: for a more realistic interaction it can be
obtained as the solution of a determined integral equation.  The fact that
our construction ensures the kernel's consistency with the vertex and hence
preservation of Ward-Takahashi identities is manifest in the Goldstone boson
nature of pion, which is preserved order-by-order and in the infinite
resummation.
 
Our explicit calculations focused primarily on flavour-nonsinglet
pseudoscalar mesons and vector mesons.  We found that a consistent,
nonperturbative dressing of the vertex and kernel changes the masses of these
mesons by $\lesssim 10$\% cf.\ the values obtained using the rainbow-ladder
truncation.  That is not the case in the pseudoscalar channel if the kernel
is dressed inconsistently.  Furthermore, $90$\% of the $\pi$-$\rho$
mass-splitting is already generated in the rainbow-ladder truncation, which
emphasises that this splitting is primarily driven by DCSB.  The
rainbow-ladder truncation is a poor approximation for flavour-singlet
pseudoscalar mesons and scalar mesons.
 
We also considered quark-quark scattering and found that, with anything but a
ladder-like vertex-consistent Bethe-Salpeter kernel, diquark bound states do
not exist in the spectrum.
 
\begin{acknowledgments} 
We are pleased to acknowledge interactions with M.~Bhagwat, A.W.~Schreiber,
P.C.~Tandy.  WD is grateful for the hospitality of the Physics Division at
Argonne National Laboratory during a visit in which part of this work was
conducted and for financial support from the Division; and CDR is likewise
grateful for the hospitality and financial support of the Special Research
Centre for the Subatomic Structure of Matter.  This work was supported by:
Adelaide University; the Australian Research Council; the Deutsche
Forschungsgemeinschaft, under contract no.\ Ro~1146/3-1; and the US
Department of Energy, Nuclear Physics Division, under contract
no.~\mbox{W-31-109-ENG-38}; and benefited from the resources of the US
National Energy Research Scientific Computing Center.
\end{acknowledgments}


\begin{thebibliography}{99}
\bibitem{fn:Eucl} We employ a Euclidean metric throughout, with: 
$\{\gamma_\mu,\gamma_\nu\} = 2\delta_{\mu\nu}$; $\gamma_\mu^\dagger = 
\gamma_\mu$; and $a \cdot b = \sum_{i=1}^4 a_i b_i$.
\bibitem{fn:confinement} Confinement is exhibited in this model, and others
in this class, via the absence of a spectral representation for coloured
$n$-point functions. This is a sufficient condition because of the associated
violation of reflection positivity, as discussed, e.g., in Sec.~6.2 of
Ref.~\cite{cdragw}, Sec.~2.2 of Ref.~\cite{revbasti} and Sec.~2.4 of
Ref.~\cite{revreinhard}.
\end{thebibliography}
\end{document}